\documentclass[12pt]{article}
\usepackage{algorithm,algpseudocode,amsmath,amssymb,amsthm,bm,graphicx,natbib,subcaption,placeins,setspace,fullpage}

\begin{document}

\title{The controlled thermodynamic integral for Bayesian model comparison}
\author{Chris J. Oates, Theodore Papamarkou, Mark Girolami}
\date{}
\maketitle

\abstract{Bayesian model comparison relies upon the model evidence, yet for many models of interest the model evidence is unavailable in closed form and must be approximated. 
Many of the estimators for evidence that have been proposed in the Monte Carlo literature suffer from high variability.
This paper considers the reduction of variance that can be achieved by exploiting control variates in this setting.
Our methodology is based on thermodynamic integration and applies whenever the gradient of both the log-likelihood and the log-prior with respect to the parameters can be efficiently evaluated.
Results obtained on regression models and popular benchmark datasets demonstrate a significant and sometimes dramatic reduction in estimator variance and provide insight into the wider applicability of control variates to Bayesian model comparison.}


{\bf Author Footnote.} 
Chris J. Oates (E-mail: c.oates@warwick.ac.uk) is Research Fellow, Theodore Papamarkou (E-mail: t.papamarkou@warwick.ac.uk) is Research Associate and Mark Girolami (E-mail: m.girolami@warwick.ac.uk) is Professor, Department of Statistics, University of Warwick, Coventry, CV4 7AL.
This work was supported by UK EPSRC EP/D002060/1, EP/J016934/1, EU Grant 259348 (Analysing and Striking the Sensitivities of Embryonal Tumours) and a Royal Society Wolfson Research Merit Award.
The authors are grateful to Christian Robert for discussions on the use of control variates in this setting.

\doublespacing

\section{Introduction}

In hypothesis-driven research we are presented with data $\bm{y}$ that is assumed to have arisen under one of two (or more) putative models $m_i$ characterised by a probability density $p(\bm{y}|m_i)$. 
Given {\it a priori} model probabilities $p(m_i)$, the data $\bm{y}$ induce {\it a posteriori} probabilities $p(m_i|\bm{y})$ that are the basis for Bayesian model comparison. 
Since any prior probability distribution gets transformed to a posterior probability distribution through consideration of the data, the transformation itself represents the evidence provided by the data \citep{Kass}. 
For the simple case of two models, this transformation follows from Bayes' rule as 
\begin{eqnarray}
\underbrace{\frac{p(m_2|\bm{y})}{p(m_1|\bm{y})}}_\text{posterior odds} = \underbrace{\frac{p(\bm{y}|m_2)}{p(\bm{y}|m_1)}}_{\text{Bayes factor }B_{21}} \times \underbrace{\frac{p(m_2)}{p(m_1)}}_\text{prior odds}.
\end{eqnarray}
Thus the influence of the data on the posterior probability distribution is captured through that Bayes factor $B_{21}$ in favour of Model 2 over Model 1.
Rearranging, we can interpret the Bayes factor as the ratio of the posterior odds to the prior odds.
A natural approach to computation of Bayes factors is to directly compute the evidence
\begin{eqnarray}
p(\bm{y}|m_i) = \int p(\bm{y}|\bm{\theta},m_i) p(\bm{\theta}|m_i) d\bm{\theta}, 
\label{ev}
\end{eqnarray}
provided by data $\bm{y}$ in favour of model $m_i$, where $\bm{\theta}$ are parameters associated with model $m_i$. 
Yet for almost all models of interest, the evidence is unavailable in closed form and must be approximated.
Numerous techniques have been proposed to approximate the model evidence (Eqn. \ref{ev}), a selection of which includes {\it path sampling} \citep{Ogata,Gelman}, {\it harmonic means} \citep{Gelfand}, {\it Chib's method} \citep{Chib}, {\it nested sampling} \citep{Skilling}, {\it particle filters} \citep{delMoral}, {\it multicanonical algorithms} \citep{Marinari,Geyer}, {\it approximate Bayesian computation} \citep{Didelot} and {\it variational approximations} \citep{Corduneanu}.
Alternatively one can directly target the Bayes factor $B_{21}$ that compares between two models.
Here too numerous methods have been proposed, including {\it importance sampling} \citep{Gelman,Chen}, {\it ratio importance sampling} \citep{Torrie}, {\it bridge sampling} \citep{Gelman,Chen}, {\it sequential Monte Carlo} \citep{Zhou}, {\it annealed importance sampling} \citep{Neal}, {\it reversible-jump Markov chain Monte Carlo} \citep[MCMC;][]{Green} and also again {\it approximate Bayesian computation} \citep{Toni}.
Recent reviews of these methodologies include \cite{Vyshemirsky,Marin,Friel3}.

Of the estimators of evidence that are based on Monte Carlo sampling, it remains the case that estimator variance can in general be extremely high.
General approaches to reduction of Monte Carlo error that have been proposed in the literature include {\it antithetic variables} \citep{Green2}, {\it control variates} and {\it Rao-Blackwellisation} \citep{Robert}, {\it Riemann sums} \citep{Philippe} and a plethora of MCMC schemes that aim to improve mixing \citep[e.g.][]{Girolami}.
These methods could all be used to reduce the variance of estimators for model evidence that are based on computing Monte Carlo expectations.
In this paper we extend the {\it zero-variance} (ZV) control variate technique, introduced in the physics literature by \cite{Assaraf}, to estimators of model evidence that are based on MCMC and {\it thermodynamic integration} \citep[TI;][]{Frenkel}.
The methodology applies whenever the gradient of the log-likelihood (and the log-prior) can be evaluated and therefore can be used ``for free'' when differential geometric sampling schemes are employed in construction of the Markov chain \citep{Papamarkou}.
Theoretical results are provided that guide maximal variance reduction in practice.
Results on popular benchmark datasets demonstrate a substantial reduction in variance compared to existing estimators and the method is shown to be exact in the special case of Bayesian linear regression.

The paper proceeds as follows:
Section \ref{back} recalls key ideas from TI and ZV that we use in our methodology.
In section \ref{methods} we derive control variates for TI and provide theoretical results that guide maximal variance reduction in practice.
Section \ref{applications} compares the proposed methodology to the state-of-the-art estimators of model evidence applied to popular benchmark datasets.
Section \ref{failure} investigates scenarios where the proposed methodology is likely to fail.
Finally section \ref{discuss} provides more general insight into the use of control variates in estimation of model evidence, drawing an important distinction between ``equilibrium'' and ``non-equilibrium'' estimators that determines whether or not control variates may be applicable.

\section{Background} \label{back}

\subsection{Thermodynamic integration}

Path sampling and the closely related technique of TI emerged from the physics community as a computational approach to compute normalising constants \citep{Gelman}.
Recent empirical investigations, including \cite{Vyshemirsky,Friel3}, have revealed that TI is among the most promising approach to estimation of model evidence.
Below we provide relevant background on TI, referring the reader to \cite{Calderhead} for a detailed discussion of implementational details.

TI targets the model evidence directly; in what follows we therefore implicitly condition upon a model $m$ and aim to compute the evidence $p(\bm{y}) = p(\bm{y}|m)$ provided by data $\bm{y}$ in favour of model $m$.
Following the presentation of \cite{Friel}, the {\it power posterior} is defined as $p(\boldsymbol{\theta} | \bm{y},t) = p(\bm{y}|\boldsymbol{\theta})^tp(\boldsymbol{\theta}) / \mathcal{Z}_t(\bm{y})$ where the normalising constant is given by $\mathcal{Z}_t(\bm{y}) = \int p(\bm{y}|\boldsymbol{\theta})^tp(\boldsymbol{\theta}) d\bm{\theta}$.
Here $t$ is known as an {\it inverse temperature} parameter and by analogy the process of increasing $t$ is known as {\it annealing}.
Note that $p(\boldsymbol{\theta} | \bm{y},t=0)$ is the density of the prior distribution, whereas $p(\boldsymbol{\theta} | \bm{y},t=1)$ is the density $p(\boldsymbol{\theta} | \bm{y})$ of the posterior distribution.
Varying $t \in (0,1)$ produces a continuous path between these two distributions and in this paper it is assumed that all intermediate distributions exist and are well-defined.
The normalising constant $\mathcal{Z}_0(\bm{y})$ is equal to one and $\mathcal{Z}_1(\bm{y})$ is equal to $p(\bm{y})$, the model evidence that we aim to estimate.

The standard thermodynamic identity is
\begin{eqnarray}
\log(p(\bm{y})) =  \int_0^1 \mathbb{E}_{\bm{\theta}|\bm{y},t} \log(p(\bm{y}|\bm{\theta})) dt
\label{ML}
\end{eqnarray}
where the expectation in the integrand is with respect to the power posterior whose density is given above.
The correctness of Eqn. \ref{ML} is established in e.g. \cite{Friel}.
In TI, this one-dimensional integral is evaluated numerically using a quadrature approximation over a discrete temperature ladder, whereas in the related approach of path sampling this integral is evaluated using MCMC.
Note that the use of quadrature methods introduces bias into the estimator of model (log-)evidence; it is therefore important to select an accurate quadrature approximation (Appendix \ref{A quadrature}).

\subsection{Control variates and the ZV technique}

Control variates are often employed when we aim to estimate, with reduced variance, the expectation $\mathbb{E}_{\pi}[g(\bm{\theta})]$ of a function $g(\bm{\theta})$ of a random variable $\bm{\theta}$ that is distributed according to a (possibly unnormalised) density $\pi(\bm{\theta})$.
In this paper we focus on real-valued $\bm{\theta} \in \Theta \subseteq \mathbb{R}^d$ and we aim to approximate
\begin{equation}
\mathbb{E}_{\pi}[g(\boldsymbol{\theta})]=
\frac{\int g(\boldsymbol{\theta})\pi(\boldsymbol{\theta})
d\boldsymbol{\theta}}
{\int \pi(\boldsymbol{\theta})d\boldsymbol{\theta}}.
\end{equation}
The generic control variate principle relies on constructing an auxiliary function $\tilde{g}(\boldsymbol{\theta})=g(\boldsymbol{\theta})+ h(\boldsymbol{\theta})$ that satisfies $\mathbb{E}_{\pi}[h(\bm{\theta})] = 0$ and so $\mathbb{E}_{\pi}[\tilde{g}(\boldsymbol{\theta})] = \mathbb{E}_{\pi}[g(\boldsymbol{\theta})]$.
Write $\mathbb{V}_\pi[g(\bm{\theta})]$ for the variance of the function $g(\bm{\theta})$ of a random variable $\bm{\theta}$ whose (unnormalised) density is $\pi(\bm{\theta})$.
In many cases it is possible to choose $h(\bm{\theta})$ such that $\mathbb{V}_{\pi}[\tilde{g}(\bm{\theta})] < \mathbb{V}_{\pi}[g(\bm{\theta})]$, leading to a reduction in Monte Carlo variance.
Intuitively, greater variance reduction can occur when $h(\bm{\theta})$ is negatively correlated with $g(\bm{\theta})$ under $\pi(\bm{\theta})$, since much of the randomness ``cancels out'' in the auxiliary function $\tilde{g}(\bm{\theta})$.
In classical literature $h(\bm{\theta})$ is formed as a sum $\phi_1 h_1(\bm{\theta}) + \dots \phi_m h_m(\bm{\theta})$ where the $h_i(\bm{\theta})$ have zero mean under $\pi(\bm{\theta})$ and are known as {\it control variates}, whilst $\phi_i$ are coefficients that must be specified.
For estimation based on Markov chains, \cite{Andradottir} proposed control variates for discrete state spaces. 
Later \cite{Mira2} extended this approach to continuous state spaces, observing that the optimal choice of $h(\bm{\theta})$ is intimately associated with the solution of the Poisson equation $h(\bm{\theta}) = \mathbb{E}_{\pi}[g(\bm{\theta})] - g(\bm{\theta})$ and proposing to solve this equation numerically.
Further work on constructing control variates for Markov chains includes \cite{Hammer} for Metropolis-Hasings chains and \cite{Dellaportas} for Gibbs samplers.

In this paper we consider the particularly tractable class of ZV control variates that are expressed as functions of the gradient $\nabla_{\bm{\theta}} \log \pi(\bm{\theta})$ of the log-target density (i.e. the score function).
More specifically, \cite{Mira} proposed to use 
\begin{eqnarray}
h(\bm{\theta}) = -\frac{1}{2} \Delta_{\bm{\theta}}[P(\bm{\theta})] + \nabla_{\bm{\theta}}[P(\bm{\theta})] \cdot \bm{z}(\bm{\theta})
\label{ZV h}
\end{eqnarray}
where the {\it trial function} $P(\bm{\theta})$ is a polynomial in $\bm{\theta}$ and
\begin{equation}
  \label{eq:auxiliary:g:poly:linear:z}
  \bm{z}(\boldsymbol{\theta})= -\frac{1}{2} \nabla_{\boldsymbol{\theta}}[\log(\pi(\boldsymbol{\theta}))]
\end{equation}
is proportional to the score function.
In this paper we adopt the convention that both $\bm{\theta}$ and $\bm{z}(\bm{\theta})$ are $d \times 1$ vectors.
The thermodynamic identity (Eqn. \ref{ML}) is based on expected values of log-likelihoods $\log(\pi(\bm{\theta}))$.
Since $\bm{z}(\bm{\theta})$ is closely related to $\log(\pi(\bm{\theta}))$, ZV control variates appear as a natural strategy to achieve variance reduction in TI.
As shown in \cite{Mira}, ZV control variates arise naturally in certain Gaussian models, leading, in some cases, to exact (i.e. deterministic) estimators that have zero variance.
Intuitively, any density $\pi(\bm{\theta})$ that approximates a Gaussian forms a suitable candidate for implementing the ZV scheme.
Theoretical conditions for asymptotic unbiasedness of ZV have been established (Appendix \ref{A ZV}).

ZV control variates are particularly tractable for two reasons:
(i) For many models of interest it is possible to obtain a closed-form expression for Eqn. \ref{ZV h}, compared to alternatives that require numerical solution of the Poisson equation;
(ii) As recently noticed by \cite{Papamarkou}, the ZV technique can be applied essentially ``for free'' inside differential-geometric MCMC sampling schemes for which the score function is a pre-requisite for sampling \citep{Girolami}.

\section{Methodology} \label{methods}

In section \ref{CVs for t} we develop a control variate scheme for the estimation of model evidence, taking TI as our base estimator whose variance we propose to reduce.
The main methodological challenge in this setting is the elicitation of both the optimal control variate coefficients $\bm{\phi}$ and the optimal temperature ladder that underlies TI.
In section \ref{elic coeff} we derive optimal expressions for both these quantities and in section \ref{compute} we describe how coefficients and temperature ladders are selected in practice.

\subsection{The controlled thermodynamic integral} \label{CVs for t}

Taking the target density $\pi(\bm{\theta})$ to be the power posterior $p(\bm{\theta}|\bm{y},t)$, it follows from Eqn. \ref{eq:auxiliary:g:poly:linear:z} that  
\begin{eqnarray}
\bm{z}(\bm{\theta}|\bm{y},t) = -\frac{t}{2}
\frac{\nabla_{\boldsymbol{\theta}}p(\bm{y}|\boldsymbol{\theta})}
{p(\bm{y}|\boldsymbol{\theta})}
- \frac{1}{2} \frac{\nabla_{\boldsymbol{\theta}}p(\boldsymbol{\theta})}
{p(\boldsymbol{\theta})}.
\label{Z path}
\end{eqnarray}
The ZV control variates (Eqn. \ref{ZV h}) are then
\begin{eqnarray}
h(\bm{\theta}|\bm{y},t) = -\frac{1}{2} \Delta_{\bm{\theta}}[P(\bm{\theta}|\bm{\phi}(\bm{y},t))] + \nabla_{\bm{\theta}}[P(\bm{\theta}|\bm{\phi}(\bm{y},t))] \cdot \bm{z}(\bm{\theta}|\bm{y},t)
\label{path improve}
\end{eqnarray}
where $\bm{z}(\bm{\theta}|\bm{y},t)$ is as defined in Eqn. \ref{Z path}.
Here the coefficients $\bm{\phi} \equiv \bm{\phi}(\bm{y},t)$ of the polynomial $P$ will in general depend on both the data $\bm{y}$ and inverse temperature $t$.
Integrating these control variates into TI, we obtain the ``controlled thermodynamic integral'' (CTI)
\begin{eqnarray}
\log(p(\bm{y})) = \int_0^1 \mathbb{E}_{\bm{\theta}|\bm{y},t} [ \log(p(\bm{y}|\bm{\theta})) + h(\bm{\theta}|\bm{y},t) ] dt.
\label{cont path samp}
\end{eqnarray}

In order to use CTI to estimate the model (log-)evidence we need to specify both (i) polynomial coefficients $\bm{\phi}(\bm{y},t)$ and (ii) an appropriate discretisation $0 = t_0 < t_1 < \dots < t_m = 1$ (the {\it temperature ladder}) of the one dimensional integral.
Specification of both polynomial coefficients and temperature ladder should be targeted at minimising the variance of CTI (see below).

\subsection{Optimal coefficients and ladders} \label{elic coeff}

We derive the jointly optimal, variance-minimising, polynomial coefficients and temperature ladder. 
For the latter, note that there is a surjective mapping from partitions $0 = t_0 < t_1 < \dots < t_m = 1$ to probability distributions on $[0,1]$ with density function $p(t)$ that is given by $\int_0^{t_i} p(s) ds = \frac{i}{m}$.
For the development below it is convenient to focus on optimising the density $p(t)$, mapping back to the temperature ladder during implementation (see section \ref{compute} below).
For clarity of the exposition write $g(\bm{\theta}) = \log (p(\bm{y}|\bm{\theta}))$ where we temporarily suppress dependence on both data $\bm{y}$ and model $m$. 
The CTI identity can be rewritten as
\begin{eqnarray}
\log(p(\bm{y})) = \int_0^1 \mathbb{E}_{\bm{\theta}|\bm{y},t}[g(\bm{\theta}) + h(\bm{\theta}|t)] dt = \mathbb{E}_{\bm{\theta},t|\bm{y}}\left[\frac{g(\bm{\theta}) + h(\bm{\theta}|t)}{p(t)}\right]
\end{eqnarray}
where the final expectation is taken with respect to the distribution with density $p(\bm{\theta},t|\bm{y}) = p(\bm{\theta}|\bm{y},t)p(t)$.
Under an approximation that Monte Carlo samples are obtained independently, so-called ``perfect transitions'', the variance of the estimator of model (log-)evidence is given by
\begin{eqnarray}
\frac{1}{N} \left\{ \int_0^1  \frac{\mathbb{E}_{\bm{\theta}|\bm{y},t} [ (g(\bm{\theta}) + h(\bm{\theta}|t))^2 ]}{p(t)} dt - [\log(p(\bm{y}))]^2 \right\}
\label{variance}
\end{eqnarray}
where $N$ is the number of Monte Carlo samples.

The optimal choice of polynomial coefficients $\bm{\phi}(t)$ and temperature ladder $p(t)$ are defined as the pair that jointly minimise Eqn. \ref{variance}. 
Specifically, we seek to minimise the Lagrangian
\begin{eqnarray}
\int_0^1  \frac{\mathbb{E}_{\bm{\theta}|\bm{y},t} [ (g(\bm{\theta}) + h(\bm{\theta}|t))^2 ]}{p(t)} dt  + \lambda \int_0^1 p(t)
\label{lagrange}
\end{eqnarray}
over $(p,\bm{\phi}):[0,1] \rightarrow \mathbb{R}^{e+1}$ where $e$ is the dimension of $\bm{\phi}$ and depends on the degree of the polynomial $P(\bm{\theta}|\bm{\phi})$ that is being employed. 
Here $\lambda$ is a Lagrange multiplier that will be used to ensure $\int p(t)dt=1$.
Below we consider degree 1 polynomials $P(\bm{\theta}|\bm{\phi}) = \bm{\theta}^T \bm{\phi}$ so that $h(\bm{\theta}|t) = \bm{\phi}(t)^T\bm{z}(\bm{\theta}|t)$ but the derivation applies analogously to higher degree polynomials, as explained in Appendix \ref{A QZV}.
The solution $(p^*,\bm{\phi}^*)$ of the Lagrangian optimisation problem (Eqn. \ref{lagrange}) is 
\begin{eqnarray}
\bm{\phi}^*(t) & = & -\mathbb{V}_{\bm{\theta}|\bm{y},t}^{-1}[\bm{z}(\boldsymbol{\theta})] \mathbb{E}_{\bm{\theta}|\bm{y},t}[g(\boldsymbol{\theta})\bm{z}(\boldsymbol{\theta})] \label{optimal phi} \\
p^*(t) & \propto & \sqrt{\mathbb{E}_{\bm{\theta}|\bm{y},t}[g(\bm{\theta})^2] - \mathbb{E}_{\bm{\theta}|\bm{y},t}[g(\bm{\theta})\bm{z}(\bm{\theta})]^T \mathbb{V}_{\bm{\theta}|\bm{y},t}[\bm{z}(\bm{\theta})]^{-1} \mathbb{E}_{\bm{\theta}|\bm{y},t}[g(\bm{\theta})\bm{z}(\bm{\theta})]} \label{optimal ladder}
\end{eqnarray}
where $\mathbb{V}_{\bm{\theta}|\bm{y},t}[\bm{z}(\boldsymbol{\theta})]$ and $\mathbb{E}_{\bm{\theta}|\bm{y},t}[g(\boldsymbol{\theta})\bm{z}(\boldsymbol{\theta})]$ denote respectively variance and cross-covariance matrices (since $\mathbb{E}_{\bm{\theta}|\bm{y},t}[\bm{z}(\bm{\theta})] = \bm{0}$).
Notice that the optimal temperature ladder for CTI is not the same as the optimal ladder for standard TI, which is given by $p^*(t) \propto \sqrt{\mathbb{E}_{\bm{\theta}|\bm{y},t}[g(\bm{\theta})^2]}$ \citep{Calderhead}.

It can be shown \citep{Rubinstein} that this choice of polynomial coefficients $\bm{\phi} = \bm{\phi}^*$ is characterised as the minimiser of the variance ratio
\begin{eqnarray}
R(t) := \frac{\mathbb{V}_{\bm{\theta}|\bm{y},t}[g(\bm{\theta}) + \bm{\phi}(t)^T \bm{z}(\bm{\theta}|t)]}{\mathbb{V}_{\bm{\theta}|\bm{y},t}[g(\bm{\theta})]} 
\label{R}
\end{eqnarray}
and at this minimum 
\begin{eqnarray}
R(t)= 1 - \text{Corr}_{\bm{\theta}|\bm{y},t}[g(\bm{\theta}),\bm{\phi}^T \bm{z}(\bm{\theta})]^2, 
\label{corr min}
\end{eqnarray}
so that greater variance reduction is expected in the case where a linear combination of the elements of the vector $\bm{z}(\bm{\theta})$ is highly correlated with the target function $g(\bm{\theta})$.

\subsection{Implementation} \label{compute}

For most models of interest both Eqn. \ref{optimal phi} and Eqn. \ref{optimal ladder} do not possess closed-form expressions and it becomes necessary to employ estimates or approximations to the optimal values.
We begin by noting that Eqn. \ref{optimal phi} actually defines the optimal, variance-minimising, coefficients independently of the choice of temperature ladder $p(t)$; this is directly verified from the Euler-Lagrange equations applied to $\bm{\phi}:[0,1] \rightarrow \mathbb{R}^e$ where $p(t)$ is held fixed.
This observation allows us to discuss these two aspects of the implementation separately:

\subsubsection{Polynomial coefficients} \label{plug in coeff}

Optimal coefficients for control variates are typically estimated based on the same sequence of MCMC samples that will subsequently be used to compute the controlled expectations \citep{Robert}.
Specifically, to estimate the optimal control variate coefficients $\bm{\phi}^*(t)$ we exploit MCMC samples to estimate both the covariance $\hat{\mathbb{V}}_{\bm{\theta}|\bm{y},t}[\bm{z}(\bm{\theta})]$ and the cross-covariance $\hat{\mathbb{E}}_{\bm{\theta}|\bm{y},t}[g(\boldsymbol{\theta})\bm{z}(\boldsymbol{\theta})]$.
These estimates are then plugged directly into Eqn. \ref{optimal phi} in order to obtain an estimate
\begin{eqnarray}
\bm{\phi}^*(t) \approx - \hat{\mathbb{V}}_{\bm{\theta}|\bm{y},t}[\bm{z}(\bm{\theta})]^{-1} \hat{\mathbb{E}}_{\bm{\theta}|\bm{y},t}[g(\boldsymbol{\theta})\bm{z}(\boldsymbol{\theta})]
\label{plug}
\end{eqnarray}
for the optimal coefficients.
Further discussion of ``plug-in'' estimators for control coefficients can be found in \cite{Dellaportas}.

\subsubsection{Temperature ladder}

For estimating the optimal temperature ladder of Eqn. \ref{optimal ladder}, one obvious numerical approach would be to firstly estimate $p^*(t)$ up to proportionality over a uniform grid $\{t_i\}$, using a preliminary MCMC run to estimate both $\mathbb{E}_{\bm{\theta}|\bm{y},t}[g(\bm{\theta})^2]$ and the covariance and cross-covariance matrices $\mathbb{V}_{\bm{\theta}|\bm{y},t}[\bm{z}(\bm{\theta})]$ and $\mathbb{E}_{\bm{\theta}|\bm{y},t}[g(\bm{\theta})\bm{z}(\bm{\theta})]$.
Then nonparametric density estimation could be applied in order to obtain an estimate for the optimal ladder $\{t_i\}$.
However this two-step procedure is computationally burdensome.
\cite{Neal2} showed that a geometric temperature ladder is optimal for annealing on the scale parameter of a Gaussian and \cite{Behrens} extended this result to target distributions of the same form as $g(\bm{\theta})$, which includes Gaussians.
In this paper we fix a quintic temperature ladder $t_i = (i/50)^5$ for use in all applications; this ladder is widely used in the TI literature and has demonstrated strong performance in empirical studies \citep[e.g.][]{Calderhead,Friel2}.
The question of how to select appropriate temperature ladders in practice is an ongoing area of research and the recent contributions of \cite{Miasojedow,Behrens,Zhou,Friel2} are compatible with our methodology.

\subsubsection{Quadrature}

The second order quadrature method of \cite{Friel2}, described in Appendix \ref{A quadrature}, requires us also to estimate the variance $\mathbb{V}_{\bm{\theta}|\bm{y},t}[\log(p(\bm{y}|\bm{\theta}))]$ at each step in the temperature ladder.
In experiments below we use ZV control variates to estimate this variance, using the identity
\begin{eqnarray}
\mathbb{V}_{\bm{\theta}|\bm{y},t}[\log(p(\bm{y}|\bm{\theta}))] = \mathbb{E}_{\bm{\theta}|\bm{y},t}  \left[\log(p(\bm{y}|\bm{\theta})) - \mathbb{E}_{\bm{\theta}|\bm{y},t} [\log(p(\bm{y}|\bm{\theta}))] \right]^2
\end{eqnarray}
and applying control variates in the estimation of each of these expectations.

\section{Applications} \label{applications}

We present several empirical studies that compare CTI to the state-of-the-art TI estimators of \cite{Friel2}.
In all applications below we base estimation on the output of a population MCMC sampler \citep{Jasra} limited to $N$ iterations at each of the 51 rungs of the temperature ladder (a total of $51\times N$ evaluations of the likelihood function).
In brief, the within-temperature proposal was provided by the {\it manifold Metropolis-adjusted Langevin algorithm} (mMALA) of \cite{Girolami}, whilst the between-temperature proposal randomly chooses a pair of (inverse) temperatures $t_i$ and $t_j$, proposing to swap their state vectors with probability given by the Metropolis-Hastings ratio \citep{Calderhead}.
To ensure fairness, the same samples were used as the basis for all estimators of model evidence, ensuring that all estimators require essentially the same amount of computation (since the score function is computed as a matter of course in mMALA).
Moreover, to explore the statistical properties of the estimators themselves, we generated 100 independent realisations of the population MCMC and thus 100 realisations of each estimator.
Full details are provided in the Supplement.

\subsection{Bayesian linear regression} 

\subsubsection{Known precision} \label{lin known}

We begin with an analytically tractable problem in Bayesian linear regression.
The (log-)likelihood function is given by 
\begin{eqnarray}
\log p(\bm{y}|\bm{X},\bm{\beta},\sigma) = -\frac{n}{2} \log(2\pi\sigma^2)  - \frac{1}{2\sigma^2} (\bm{y}-\bm{X}\bm{\beta})^T(\bm{y}-\bm{X}\bm{\beta})
\label{like1}
\end{eqnarray}
where $\bm{y}$ is $n \times 1$, $\bm{X}$ is $n \times d$ and $\beta$ is $d \times 1$.
In simulations below we took $\sigma = 1$, $d = 3$, $\bm{\beta} = [0,1,2]$. 
The design matrix $\bm{X}$ was populated with $n = 100$ rows by drawing each entry independently from the standard normal distribution and then data $\bm{y}$ were generated from $N(\bm{X}\bm{\beta},\sigma^2 \bm{I}_{n \times n})$; both $\bm{X}$ and $\bm{y}$ were then fixed for all experiments below.
From the Bayesian perspective we take a conjugate prior $\bm{\beta} \sim N(\bm{0},\zeta^2\bm{I}_{d \times d})$ with $\zeta = 1$.
In this section we assume $\sigma$ is fixed and known, but we relax this assumption in the next section.
Thus the unknown model parameters here are $\bm{\theta} = \bm{\beta} \in \mathbb{R}^d$ and we aim to compute the evidence $p(\bm{y})$ by marginalising over these parameters.
This example is an ideal benchmark since it is permissible to obtain many relevant quantities in closed form; see Appendix \ref{lin known formulae} for full details.

Before applying CTI we are required to check that the sufficient conditions for the unbiasedness of ZV estimators are satisfied (see Appendix \ref{A ZV}).
This amounts to noticing that the tails of the power posterior $p(\bm{\beta}|\bm{y},t)$ decay exponentially in $\bm{\beta}$ (Appendix \ref{lin known formulae}).
Using the plug-in estimates (Eqn. \ref{plug}) we obtain estimates for the optimal coefficients $\bm{\phi}^*$, that are shown in SFig. \ref{coefficients}. 
For degree 2 polynomials we see that the plug-in estimator is deterministic.
Indeed, by direct calculation we see that $\bm{z}(\bm{\beta}|\bm{y},t)$ is an invertible affine transformation of the parameter vector
\begin{eqnarray}
\bm{z}(\bm{\beta}|\bm{y},t) = -\frac{t}{2\sigma^2} \bm{X}^T\bm{y} + \frac{1}{2} \bm{\Sigma}(t)^{-1} \bm{\beta}
\label{CVs1}
\end{eqnarray}
where $\bm{\Sigma}(t) = ( \frac{t}{\sigma^2}\bm{X}^T\bm{X} + \frac{1}{\zeta^2}\bm{I})^{-1}$.
This allows us to intuit that CTI based on degree 2 polynomials will produce an {\it exact} estimate of the (log-)evidence (up to quadrature error), as we explain below.
Indeed, by another invertible affine transformation we can map $\bm{z}(\bm{\beta}|\bm{y},t) \mapsto \bm{y} - \bm{X} \bm{\beta}$ which, when multiplied by the polynomial $P(\bm{\beta}|\bm{\phi}) = (\bm{y} - \bm{X}\bm{\beta})^T$ produces a quantity $(\bm{y} - \bm{X} \bm{\beta})^T(\bm{y} - \bm{X} \bm{\beta})$ that is perfectly correlated with the log-likelihood under the power posterior.
It then follows from Eqn. \ref{R} that CTI will possess zero variance.
This argument is made rigorous in the Supplement.

\begin{figure}[t]
\centering
\includegraphics[width = .45\textwidth]{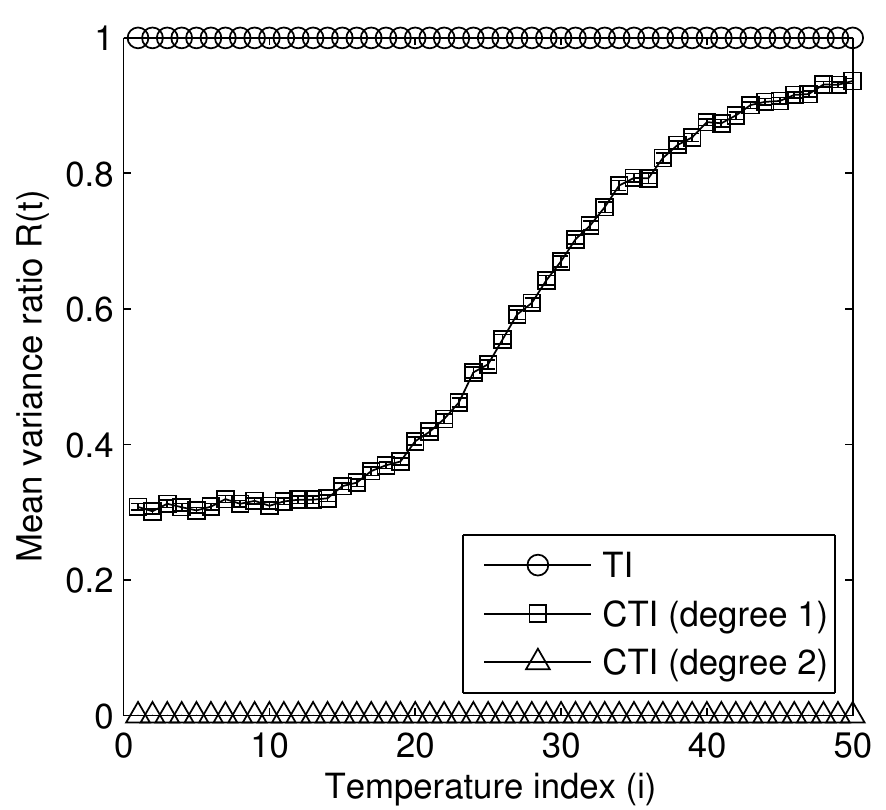} 
\caption{Bayesian linear regression, known precision. 
[Here we plot the mean variance ratio $R(t)$ computed over 100 independent runs of population MCMC using $N=1000$ samples. Error bars show standard error of these mean estimates. 
The x-axis records the index $i$ corresponding to (inverse) temperature $t_i = (i/50)^5$.]
}
\label{lin ebar1}
\end{figure}

In SFig. \ref{integrands} we plot 100 independent estimates of the integrand $\mathbb{E}_{\bm{\beta}|\bm{y},t}[g(\bm{\beta})]$ at each of the 51 temperatures in the ladder for polynomial trial functions of degree 0 (i.e. standard TI), 1 and 2.
It is apparent that estimator variance is greatest at lower values of $t$; this motivates the heavily skewed temperature ladder used by ourselves and others, as we wish to target our computational effort on this high-variance region.
We quantify the reduction in estimator variance at an (inverse) temperature $t$ using the variance ratio $R(t)$ as estimated from the MCMC samples.
Fig. \ref{lin ebar1} shows that degree 1 polynomials achieve (on average) variance reduction at all temperatures, with the greatest reduction occurring in the region where $t$ is small. 
This is encouraging as the region where $t$ is small is most important for variance reduction of TI, as discussed above.
For degree 2 polynomials we have $R(t) = 0$ for all $t$, which recapitulates the exactness of the CTI estimator in this example.

Finally we explore the quality of the estimators of model evidence themselves. 
For this model the (log-)evidence is available in closed form (Appendix \ref{lin known formulae}) and this allows us to compute the mean square error (MSE) over all 100 independent realisations of each estimator.
Results, shown in Table \ref{regression tab}, demonstrate that CTI with degree 2 polynomials achieves a 2-fold reduction in MSE compared to standard TI when both estimators are based on first order quadrature.
However, first order quadrature is known to lead to significant estimator bias \citep{Friel2} and when estimators are based instead on more accurate second order quadrature, CTI is seen to be approximately $10,000 \times$ better that TI in terms of MSE; a dramatic difference.
We also compared TI approaches against annealed importance sampling \citep[AIS;][]{Neal}, as described in the Supplement.
In this case CTI (degree 2) is over $10,000 \times$ more accurate compared to AIS (SFig. \ref{var reduce}).

\begin{figure}[t]
\centering
\begin{subfigure}[]{0.45\textwidth}
\includegraphics[width = \textwidth]{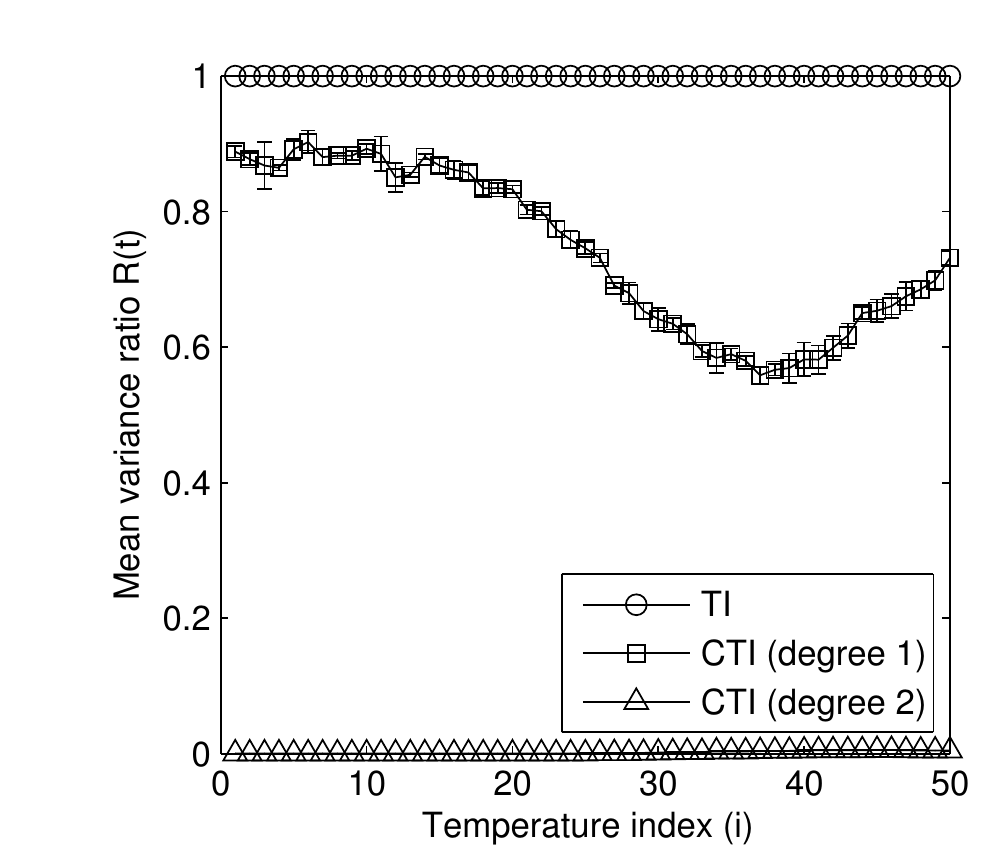}
\caption{Model 1}
\end{subfigure}
\begin{subfigure}[]{0.45\textwidth}
\includegraphics[width = \textwidth]{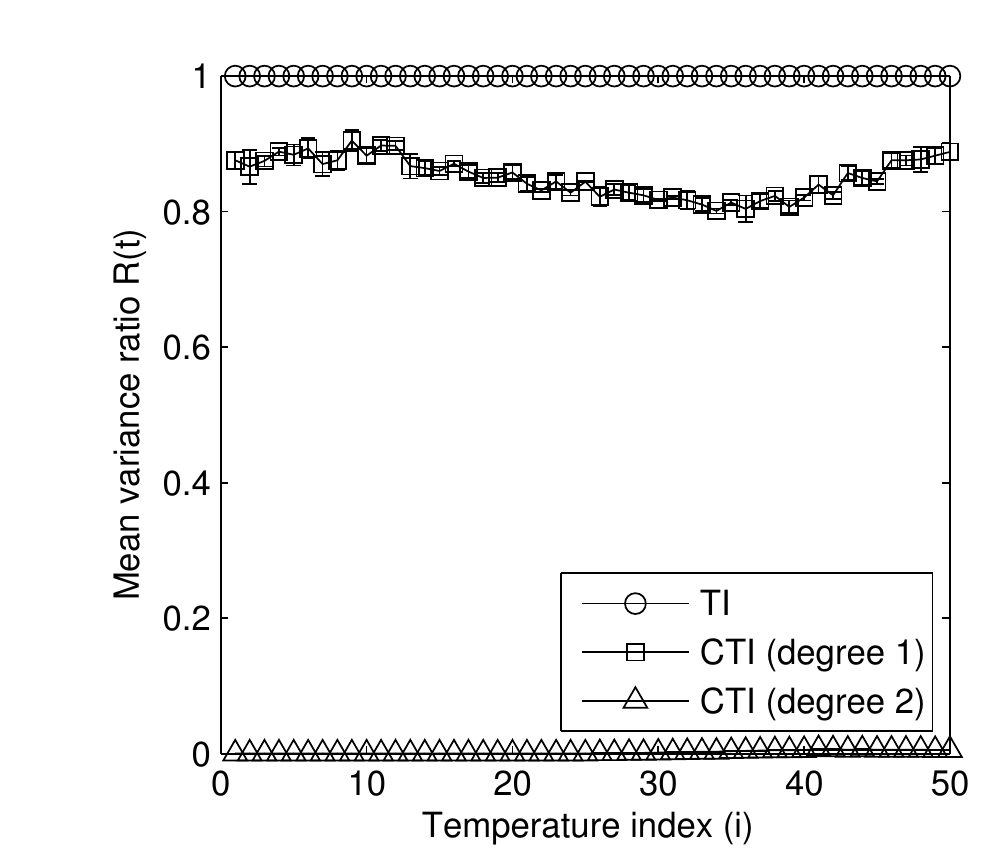}
\caption{Model 2}
\end{subfigure}
\caption{Bayesian linear regression, unknown precision.
[Here we plot the mean variance ratio $R(t)$ computed over 100 independent runs of population MCMC using $N=1000$ samples. Error bars show standard error of these mean estimates. 
The x-axis records the index $i$ corresponding to (inverse) temperature $t_i = (i/50)^5$.]
}
\label{lin ebar2}
\end{figure}

\subsubsection{Unknown precision (Radiata Pine)} \label{lin unknown}

We now relax the assumption of known precision $\tau = 1 / \sigma^2$; we will see that in these circumstances CTI is no longer exact.
Specifically we consider data from \cite{Williams} on $n=42$ specimens of {\it radiata} pine.
This dataset is well known in the multivariate statistics literature and was recently used by \cite{Friel3,Friel2} in order to benchmark estimators of model evidence.
Data consist of the maximum compression strength parallel to the grain $y_i$ as a function of density $x_i$ and density adjusted for resin content $z_i$.
It is wished to determine whether the density or resin-adjusted density is a better predictor of compression strength parallel to the grain.
Following \cite{Friel2} we consider Bayesian model comparison between a pair of competing models:
\begin{eqnarray}
\text{Model 1:} & \; \; \; y_i = \alpha + \beta(x_i - \bar{x}) + \epsilon_i, \; \; \; \epsilon_i \sim N(0,\tau^{-1}) \\
\text{Model 2:} & \; \; \; y_i = \gamma + \delta(z_i - \bar{z}) + \eta_i, \; \; \; \eta_i \sim N(0,\lambda^{-1}) 
\end{eqnarray}
Here $\bar{x}$ and $\bar{z}$ are the sample means of the $x_i$ and $z_i$ respectively.
The priors for $(\alpha,\beta)$ and $(\gamma,\delta)$ are both Gaussian with common mean $\bm{B}_0 = [3000,185]^T$ and precisions $\tau \bm{Q}_0$, $\lambda \bm{Q}_0$ where $\bm{Q}_0 = \text{diag}(0.06,6)$.
Both $\tau$ and $\lambda$ were assigned gamma priors with shape $6$ and rate $4 \times 300^2$.
To compare between these models we consider estimates for the log-Bayes factor $\log(B_{21})$ that are obtained as the difference between independent estimates for the log-evidence of each model.

This application is interesting for two reasons: Firstly, one can directly calculate the Bayes factor for this example as $B_{21} = 8.7086$, so that we have a gold standard performance benchmark.
Secondly, when the precision $\tau$ (or $\lambda$) is unknown, ZV methods are no longer exact. 
We therefore have an opportunity to assess the performance of CTI in a non-trivial setting. 

Formulae in Appendix \ref{lin unknown formulae} demonstrate that the sufficient condition for unbiasedness of ZV methods is satisfied.
Results in Fig. \ref{lin ebar2} show that CTI (degree 1) achieves a modest reduction in variance across temperatures $t$, whereas CTI (degree 2) achieves a massive variance reduction.
Computing the MSE relative to the true Bayes factor we see that CTI (degree 2) is over $500 \times$ more accurate compared to TI, though the variance of the estimator is not identically equal to zero in this case (Table \ref{regression tab}).
As before, MSE is further reduced as a result of applying second order quadrature.
AIS performed slightly worse than the methods based on TI in this example (SFig. \ref{app2 BF}).

\begin{table}[t]
\scriptsize
\centering
\begin{tabular}{c|c|c|c|c|c} \hline
{\bf Precision} & $N$ & {\bf deg}$(P)$ & {\bf Quadr.} & {\bf M.S.E.} & {\bf S.E.} \\ \hline \hline
(a) Known & 1e3 & 0 & 1 & 2.3e-2  & 2.9e-3 \\ \cline{4-6}
& &  & 2 &  2.1e-2  & 2.5e-3 \\ \cline{3-6}
& & 1 & 1 & 1.8e-2 &  2.1e-3 \\  \cline{4-6}
& &  & 2 &  2.0e-2 &  2.5e-3 \\ \cline{3-6}
& & 2 & 1 & 1.2e-2 &  0 \\  \cline{4-6}
& &  & 2 & {\bf 2.2e-6} & {\bf 1.6e-7} \\ \cline{2-6}
& 5e3 & 0 & 1 & 5.2e-3  & 7.2e-4 \\ \cline{4-6}
& &  & 2 & 4.0e-3 & 5.9e-4 \\ \cline{3-6}
& & 1 & 1 & 3.8e-3 & 6.0e-4 \\ \cline{4-6}
& &  & 2 & 3.4e-3 & 5.3e-4 \\ \cline{3-6}
& & 2 & 1 & 1.2e-3 & 0 \\ \cline{4-6}
& &  & 2 & {\bf 2.0e-7} & {\bf 2.7e-8} \\ \hline
(b) Unknown & 1e3 & 0 & 1 & 7.9e-3  & 1.2e-3 \\ \cline{4-6}
& &  & 2 &  7.7e-3   & 1.1e-3 \\ \cline{3-6}
& & 1 & 1 & 7.8e-3 & 1.0e-3 \\ \cline{4-6}
& &  & 2 &  7.6e-3 & 1.0e-3 \\ \cline{3-6}
& & 2 & 1 & 1.4e-5 & 2.0e-6 \\  \cline{4-6}
& &  & 2 & {\bf 1.3e-5} & {\bf 1.6e-6}  \\ \cline{2-6}
& 5e3 & 0 & 1 & 1.4e-3  & 1.8e-4 \\ \cline{4-6}
& &  & 2 &  1.3e-3 & 1.8e-4 \\ \cline{3-6}
& & 1 & 1 & 1.4e-3 & 2.0e-4 \\ \cline{4-6}
& &  & 2 &  1.4e-3 & 2.0e-4 \\ \cline{3-6}
& & 2 & 1 & 2.4e-6 & 3.0e-7 \\ \cline{4-6}
& &  & 2 & {\bf 1.5e-6} & {\bf 2.0e-7} \\ \hline
\end{tabular}
\caption{Bayesian linear regression with (a) known precision and (b) unknown precision. Mean square error (MSE) for estimates of the log-evidence in (a) and the Bayes factor in (b), based on 100 independent runs of population MCMC, along with estimates for standard errors (SE). 
dim$(P)$ is the dimension of the ZV polynomial $P(\bm{\theta})$, with 0 denoting standard TI.
Quadr. is the order of numerical quadrature scheme.
$N$ is the number of MCMC iterations.
}
\label{regression tab}
\end{table}

\subsection{Bayesian logistic regression (Pima Indians)} \label{logit}

Here we examine data that contains instances of diabetes and a range of possible diabetes indicators for $n = 532$ women who were at least 21 years old, of Pima Indian heritage and living near Phoenix, Arizona.
This dataset is frequently used as a benchmark for supervised learning methods \citep[e.g.][]{Marin}.
\cite{Friel2} considered seven predictors of diabetes recorded for this group; number of pregnancies (NP); plasma glucose concentration (PGC); diastolic blood pressure (BP); triceps skin fold thickness (TST); body mass index (BMI); diabetes pedigree function (DP) and age (AGE). 
Diabetes incidence $y_i$ in person $i$ is modelled by the binomial likelihood
\begin{eqnarray}
p(\bm{y}|\bm{\beta}) = \prod_{i=1}^n p_i^{y_i} (1-p_i)^{1-y_i}, 
\end{eqnarray}
where the probability of incidence $p_i$ for person $i$ is related to the covariates $\bm{x}_{i,\bullet} = (1,x_{i,1},\dots,x_{i,d})^T$ and the parameters $\bm{\beta} = (\beta_0,\beta_1,\dots,\beta_d)$ by 
\begin{eqnarray}
\text{logit}(p_i) = \log\left(\frac{p_i}{1-p_i}\right) = \bm{x}_{i,\bullet}\bm{\beta}.
\end{eqnarray}
Bayesian model comparison is desired to be performed between the two candidate models
\begin{eqnarray}
\text{Model 1:} & \; \; \; \text{logit}(p) = \beta_0 + \beta_1 \text{NP} + \beta_2\text{PGC} + \beta_3 \text{BMI} + \beta_4 \text{DP} \\
\text{Model 2:} & \; \; \; \text{logit}(p) = \beta_0 + \beta_1 \text{NP} + \beta_2\text{PGC} + \beta_3 \text{BMI} + \beta_4 \text{DP} + \beta_5 \text{AGE}
\end{eqnarray}
subject to the prior belief $\bm{\beta} \sim N(\bm{0},\tau^{-1}\bm{I})$. 
Following \cite{Friel3} we set $\tau = 0.01$.

\begin{figure*}[t]
\centering
\begin{subfigure}[]{0.45\textwidth}
\includegraphics[width = \textwidth]{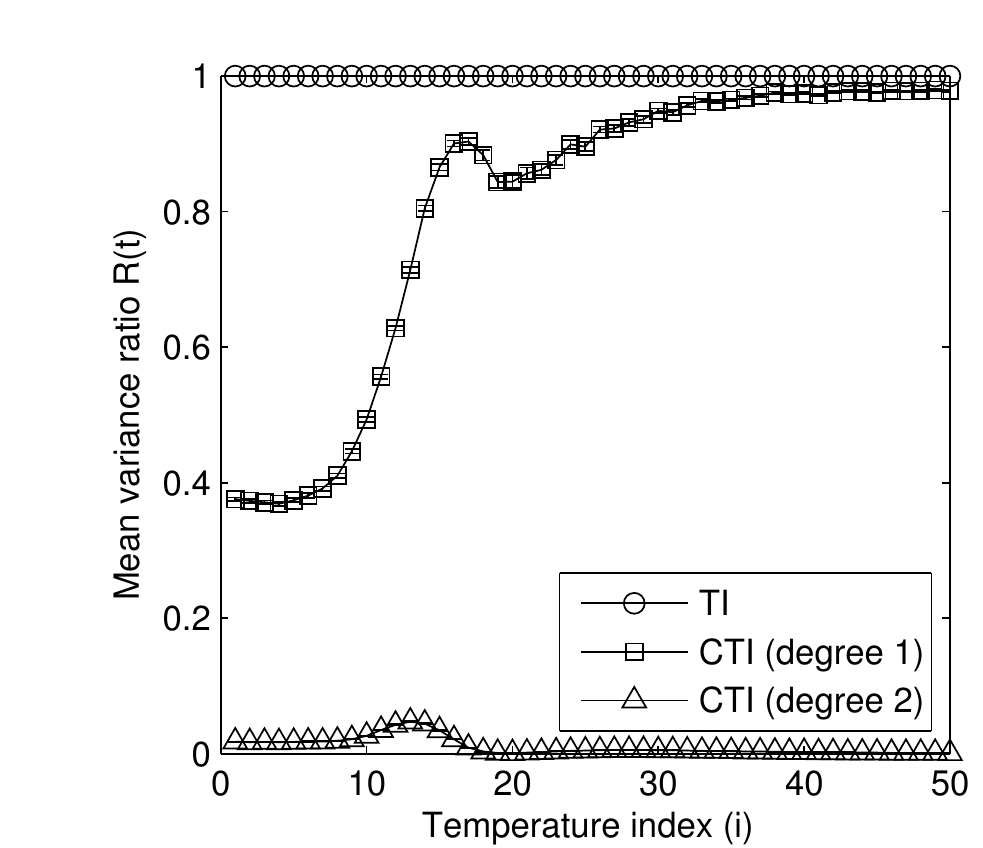}
\caption{Model 1}
\end{subfigure}
\begin{subfigure}[]{0.45\textwidth}
\includegraphics[width = \textwidth]{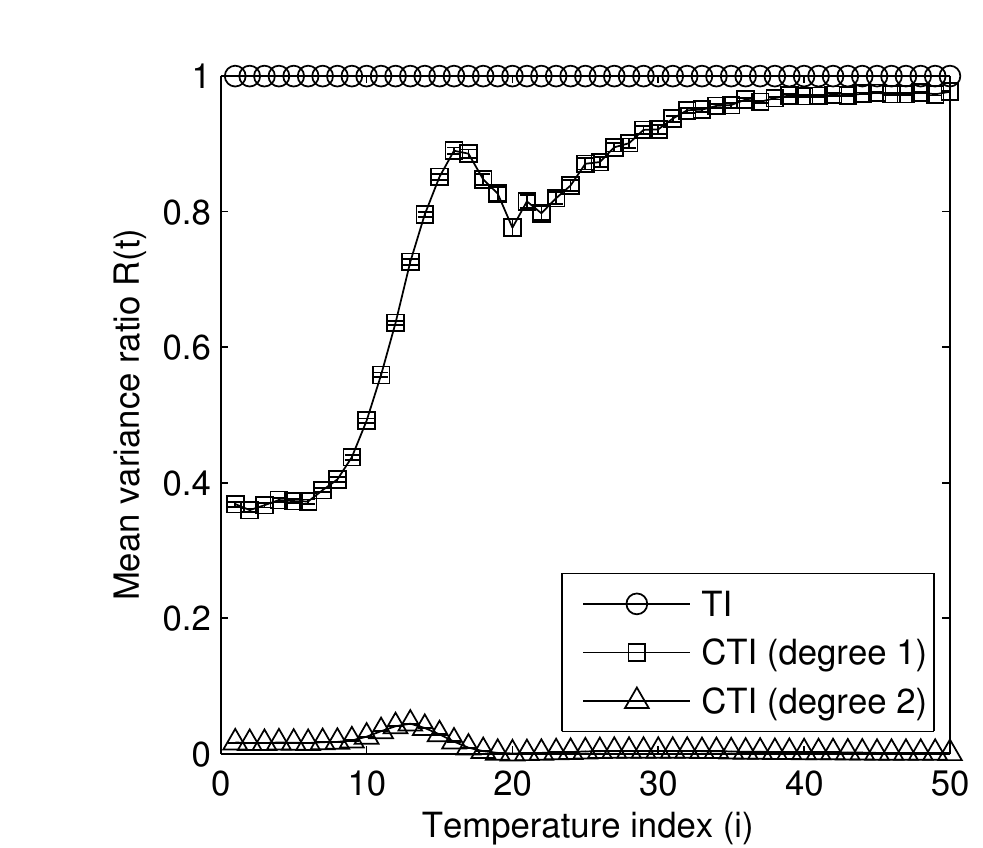}
\caption{Model 2}
\end{subfigure}
\caption{Bayesian logistic regression.
[Here we plot the mean variance ratio $R(t)$ computed over 100 independent runs of population MCMC using $N=1000$ samples. Error bars show standard error of these mean estimates. 
The x-axis records the index $i$ corresponding to (inverse) temperature $t_i = (i/50)^5$.]
}
\label{pimaVR}
\end{figure*}

The unbiasedness criterion in Appendix \ref{A ZV} is seen to be satisfied and we have
\begin{eqnarray}
\bm{z}(\bm{\beta}|\bm{y},t) = -\frac{t}{2}\bm{X}^T(\bm{y}-\bm{p}) + \frac{\tau\bm{\beta}}{2}
\end{eqnarray}
where the $i$th row of $\bm{X}$ is $\bm{x}_{i,\bullet}$.
In Fig. \ref{pimaVR} we see that degree 1 ZV methods achieve a greater variance reduction at smaller $t$, but moreover we see that degree 2 ZV methods continue to achieve a substantial variance reduction at all temperatures.
In Table \ref{app3 table} we display the mean of each estimator $\hat{B}_{21}$, computed over all 100 independent runs of population MCMC, together with the standard deviation of this collection of estimates.
We see that this variance reduction transfers to estimates of the Bayes factor themselves, where the standard deviation of the CTI estimators is approximately $20\times$ lower compared to estimators based on TI.
Although no exact expression is available for $B_{21}$, \cite{Friel2} computed the log-evidence for both models using an extended run of 2,000 temperatures and $N=20,000$ iterations within standard TI.
Their estimates were $-257.2342$ and $-259.8519$ respectively for Models 1 and 2, corresponding to an estimate of the Bayes factor of $B_{21} = -2.6177$.
This estimate, obtained at considerable computational expense, closely matches the estimates obtained by CTI (degree 2), which is based on $800 \times$ fewer evaluations of the likelihood function.
AIS performs comparably with standard TI in this example (SFig. \ref{BF logistic}).

\begin{table}[t]
\scriptsize
\centering
\begin{tabular}{c|c|c|c|c|c} \hline
{\bf Model} & $N$ & {\bf deg}$(P)$ & {\bf Quadr.} & {\bf Mean B.F.} & {\bf S.D.} \\ \hline \hline
(a) Logistic & 1e3 & 0 & 1 & -2.59 & 0.74 \\ \cline{4-6}
regression & &  & 2 &  -2.58 & 0.73 \\ \cline{3-6}
& & 1 & 1 & -2.44 & 0.70 \\ \cline{4-6}
& &  & 2 &  -2.42 & 0.69 \\ \cline{3-6}
& & 2 & 1 & -2.62 & 0.050 \\ \cline{4-6}
& &  & 2 & {\bf -2.61} & {\bf 0.044} \\ \cline{2-6}
& 5e3 & 0 & 1 & -2.62 & 0.35 \\ \cline{4-6}
& &  & 2 & -2.60 & 0.34 \\ \cline{3-6}
& & 1 & 1 & -2.58 & 0.35 \\ \cline{4-6}
& &  & 2 & -2.56 & 0.34 \\ \cline{3-6}
& & 2 & 1 & -2.64 & 0.016 \\ \cline{4-6}
& &  & 2 & {\bf -2.62} & {\bf 0.016} \\ \cline{1-6}
(b) Nonlinear & 1e3 & 0 & 1 & -3.75 & 0.31 \\ \cline{4-6}
ODEs & &  & 2 &  -3.74 & 0.31 \\ \cline{3-6}
& & 1 & 1 & -3.69 & 0.31 \\ \cline{4-6}
& &  & 2 &  -3.69 & 0.31 \\ \cline{3-6}
& & 2 & 1 & -3.57 & 0.27 \\ \cline{4-6}
& &  & 2 & {\bf -3.56} & {\bf 0.27} \\ \hline
\end{tabular}
\caption{Estimates of the log-Bayes factor $B_{21}$, based on 100 independent runs of population MCMC.
(a) Bayesian logistic regression. The actual Bayes factor, as computed by \cite{Friel2}, is $B_{21} = -2.6177$.
(b) Nonlinear ODEs: Estimates of the log-Bayes factor $B_{12}$, based on 10 independent runs of population MCMC.
dim$(P)$ is the dimension of the ZV polynomial $P(\bm{\theta})$, with 0 denoting standard TI.
Quadr. is the order of numerical quadrature scheme.
$N$ is the number of MCMC iterations.
Mean BF and SD are the mean and standard deviation of the estimated Bayes factors.
}
\label{app3 table}
\end{table}

\FloatBarrier

\section{Limitations of CTI} \label{failure}

We have demonstrated, using standard benchmark datasets, that CTI is well-suited to Bayesian model comparison between regression models. 
Regression analyses continue to be widely applicable in disciplines such as econometrics, epidemiology, political science, psychology and sociology, so that these findings have significant implications.
Nevertheless in many disciplines such as engineering, geophysics and systems biology, statistical models are significantly more complex, often based on a mechanistic understanding of the underlying process.
Below we provide such an example and find that CTI offers little improvement over TI; this allows us to explore the limitations of our approach and, conversely, to understand in what circumstances it is likely to be successful.

\subsection{A negative example (Goodwin Oscillator)} \label{odes}

We consider nonlinear dynamical systems of the form
\begin{eqnarray}
\frac{d \bm{x}}{ds} = \bm{f}(\bm{x},s;\bm{\theta}), \; \; \; \bm{x}(0) = \bm{x}_0.
\label{ODEs}
\end{eqnarray}
We assume only a subset of the variables are observed under noise, so that $\bm{x} = [\bm{x}_a,\bm{x}_b]$ and $\bm{y}$ is a $d$ by $n$ matrix of observations of the variables $\bm{x}_a$.
Model comparison for systems specified by nonlinear differential equations is known to be profoundly challenging \citep{Calderhead3}.

Write $s_1<s_2<\dots<s_n$ for the times at which observations are obtained, such that $\bm{y}(s_j) = \bm{y}_{\bullet,j}$.
We consider a Gaussian observation process with likelihood
\begin{eqnarray}
p(\bm{y}|\bm{\theta},\bm{x}_0,\sigma) = \prod_{j=1}^n \mathcal{N}(\bm{y}(s_j)|\bm{x}_a(s_j;\bm{\theta},\bm{x}_0),\sigma^2\bm{I})
\end{eqnarray}
where $\bm{x}_a(s_j;\bm{\theta},\bm{x}_0)$ denotes the solution of the system in Eqn. \ref{ODEs}.
For the Gaussian observation model it can be shown that a sufficient condition for unbiasedness of ZV is that the parameter prior density $p(\bm{\theta})$ vanishes faster than $r^{d+k-2}$ when $r = \|\bm{\theta}\|_1 \rightarrow \infty$.
Here $d = \dim\Theta$ and $k = 1$ is the degree of the polynomial that is being employed (see Appendix \ref{A ZV}).

\begin{figure*}[t]
\centering
\begin{subfigure}[]{0.45\textwidth}
\includegraphics[width = \textwidth]{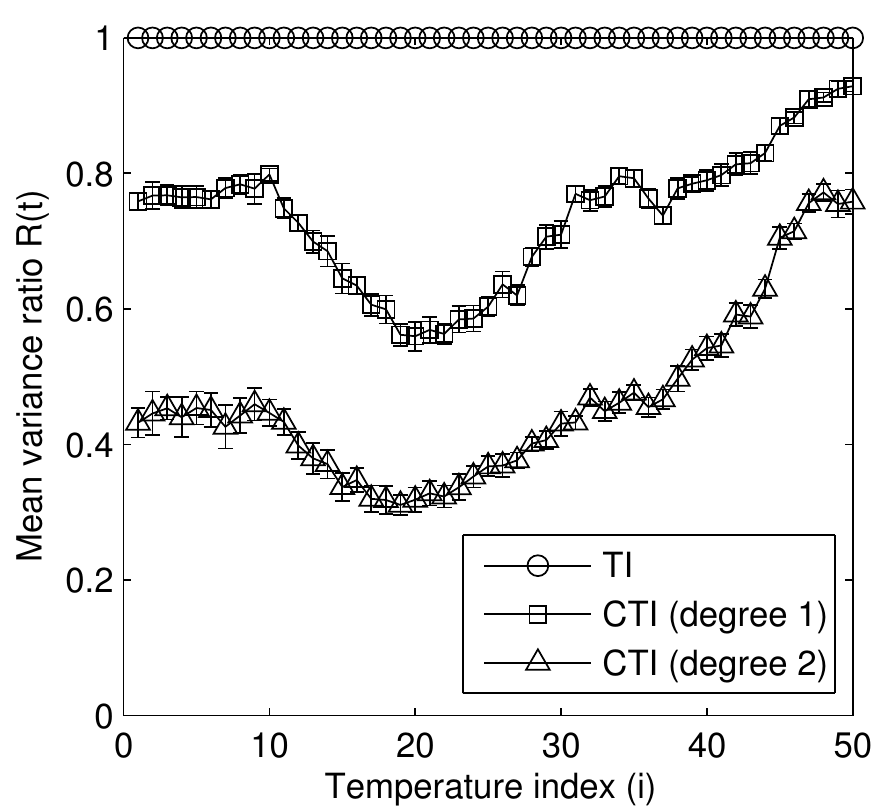}
\caption{Model 1}
\end{subfigure}
\begin{subfigure}[]{0.45\textwidth}
\includegraphics[width = \textwidth]{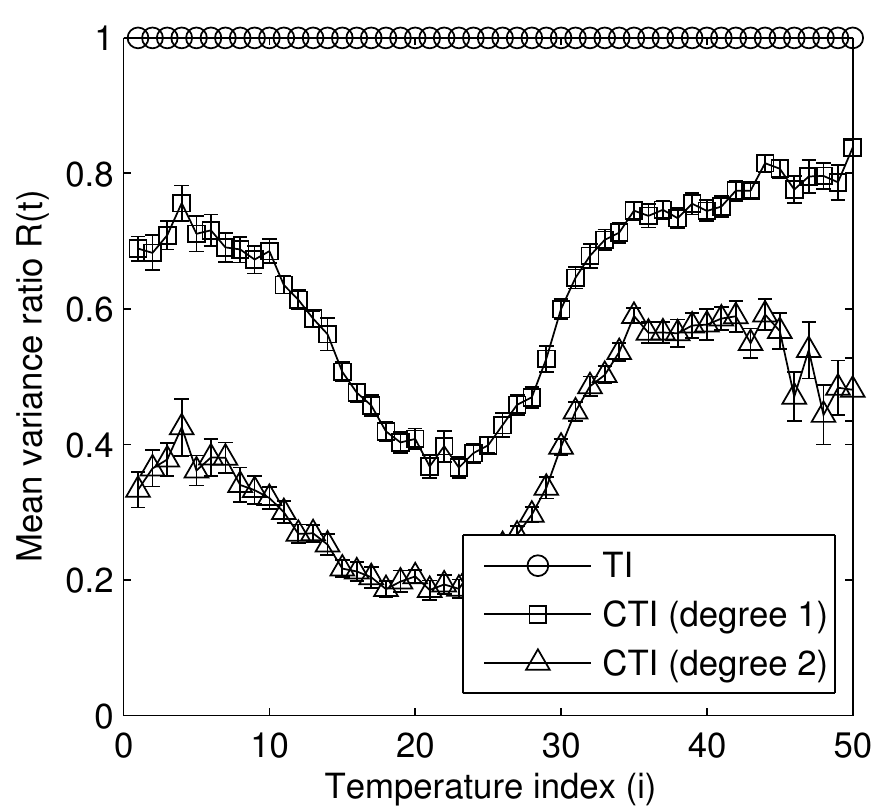}
\caption{Model 2}
\end{subfigure}
\caption{Nonlinear ODEs.
[Here we plot the mean variance ratio $R(t)$ computed over 10 independent runs of population MCMC using $N=1000$ samples. Error bars show standard error of these mean estimates. 
The x-axis records the index $i$ corresponding to (inverse) temperature $t_i = (i/50)^5$.]
}
\label{ode var}
\end{figure*}

Assuming the sufficient condition for ZV is satisfied, we have
\begin{eqnarray}
z_i(\bm{\theta}) = -\frac{t}{2\sigma^2} \sum_{j=1}^n \bm{S}_{j,1:\dim \bm{x}_a}^i (\bm{y}(s_j) - \bm{x}_a(s_j;\bm{\theta},\bm{x}_0)) -\frac{1}{2} \nabla_{\bm{\theta}} \log(p(\bm{\theta}))
\label{ODE CVs}
\end{eqnarray}
where $\bm{S}^i$ is a matrix of {\it sensitivities} with entries $S_{j,k}^i = \frac{\partial x_k}{\partial \theta_i}(s_j)$.
Note that in Eqn. \ref{ODE CVs}, $\bm{S}_{j,k}^k$ ranges over indices $1 \leq k \leq \dim \bm{x}_a$ corresponding only to the observed variables.
In general the sensitivities $\bm{S}^i$ will be unavailable in closed form, but may be computed numerically by augmenting the system of ordinary differential equations (ODEs) in Eqn. \ref{ODEs} as described in Appendix \ref{ap ode}.
Indeed, these sensitivities are already computed when differential-geometric sampling schemes are employed, so that the evaluation of Eqn. \ref{ODE CVs} incurs negligible computational cost.

\begin{figure*}[t]
\centering
\begin{subfigure}[]{0.3\textwidth}
\includegraphics[width = \textwidth]{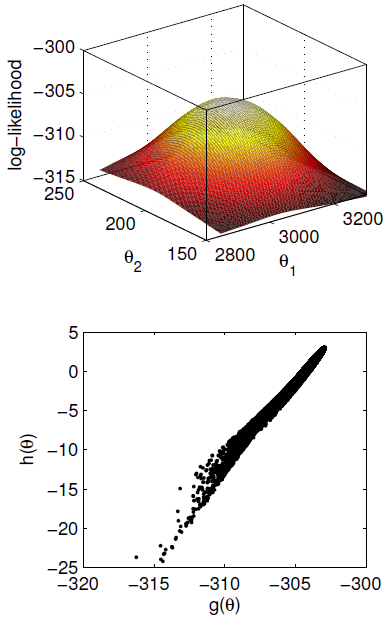}
\caption{Radiata Pine}
\end{subfigure}
\begin{subfigure}[]{0.31\textwidth}
\includegraphics[width = \textwidth]{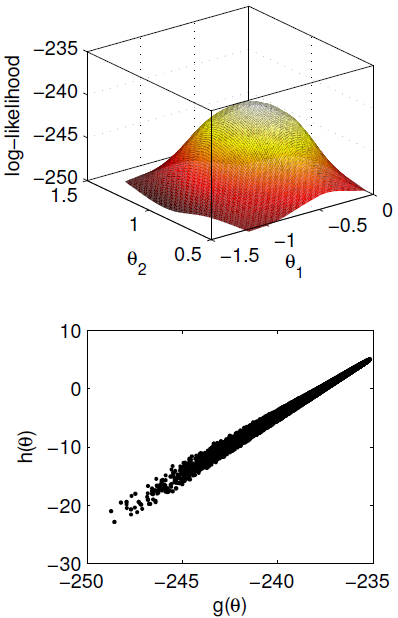}
\caption{Pima Indians}
\end{subfigure}
\begin{subfigure}[]{0.3\textwidth}
\includegraphics[width = \textwidth]{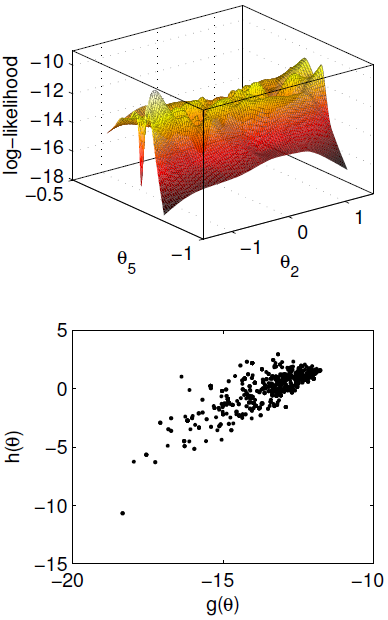}
\caption{Goodwin Oscillator}
\end{subfigure}
\caption{Comparing the likelihood surfaces and canonical correlations of different models. 
[Log-likelihood surfaces (top) for the (a) Radiata Pine and (b) Pima Indians examples can be well-approximated by a Gaussian and induce strong canonical correlation (bottom) between the (degree 2) control variates $h(\bm{\theta})$ and the log-likelihood $g(\bm{\theta})$ in the posterior. 
On the other hand, the log-likelihood surface for (c) Goodwin Oscillator is highly multi-modal and there is much weaker canonical correlation between the control variates and the log-likelihood.]}
\label{surfs}
\end{figure*}

We focus on a dynamical model of oscillatory enzymatic control due to \citep{Goodwin}, that was recently considered in the context of Bayesian model comparison by \cite{Calderhead}. 
This kinetic model, specified by a system of $g$ ODEs, describes how a negative feedback loop between protein expression and mRNA transcription can induce oscillatory dynamics as experimentally observed in circadian regulation \citep{Locke}.
A full specification is provided in Appendix \ref{ap ode}.
As shown in \cite{Calderhead}, the {\it Goodwin oscillator} induces a highly multi-modal posterior distribution that renders estimation of the model evidence extremely challenging.
We consider Bayesian comparison of two models; a simple model with one intermediate protein species ($g=3$) and a more complex model with two intermediate protein species ($g=4$).

Fig. \ref{ode var} demonstrates that in this extremely challenging example the benefits of control variate schemes that we have previously observed are heavily reduced. 
Since the variance ratio $R(t)$ is related to the canonical correlation between control variates and the log-likelihood under the power posterior (Eqn. \ref{corr min}), we hypothesise that the extreme multi-modality of the power posterior distribution is limiting the extent to which strong canonical correlation can be achieved.
This is confirmed in Fig. \ref{surfs} where we plot values of the target function $g(\bm{\theta})$ against the control variates $h(\bm{\theta})$ that are obtained from MCMC sampling in the posterior.
We observe much reduced correlation in the case of the Goodwin oscillator that is a consequence of the complex nature of the likelihood surface.
Turning to the Bayes factor itself, in Table \ref{app3 table} we display the mean of each estimator of the Bayes factor, together with the standard deviation of this collection of estimates.
We find that CTI (degree 1) provides negligible reduction in variance and CTI (degree 2) provides an insignificant $15 \%$ reduction in variance.

In this example AIS consistently produced lower estimates for Bayes factors (SFig. \ref{ode scatter}).
This likely reflects the low number $N$ of Monte Carlo iterations that are characteristic of such computationally demanding applications.

\section{Discussion} \label{discuss}

To the best of our knowledge this is the first paper to consider the use of control variates for the purpose of Bayesian model comparison.
Motivated by previous empirical studies, we focussed on TI estimators for the model (log-)evidence.
However, in general, control variate techniques could be leveraged in Bayesian model comparison whenever estimators of the evidence (or Bayes factors) take the form of a Monte Carlo expectation.
General control variate schemes for MCMC rely on the fact that the expectation of the control variates along the MCMC sample path will be approximately zero.
We thus draw a distinction between ``equilibrium'' Monte Carlo estimators for the model evidence, such as TI and path sampling, that require the underlying Markov chain to have converged, and ``non-equilibrium'' estimators such as AIS and sequential Monte Carlo that do not require convergence.
The former class are amenable to existing control variate schemes whereas the latter are not.
This motivates the ``equilibration'' of these non-equilibrium estimators.

Given its close connection with TI \citep{Gelman}, we considered whether an equilibrated version of AIS, that jointly samples from all rungs of the temperature ladder at once, would benefit from application of ZV control variates.
In contrast to CTI, the controlled AIS estimator (CAIS) demonstrated an {\it increase} in variance compared to standard AIS.
Full details are provided in the Supplement, in addition to results on each of the applications considered in this paper.
To understand these counter-intuitive results, notice that control variates must be constructed simultaneously over all $m$ rungs of the temperature ladder, so that for degree 1 polynomials we have to jointly estimate $m d$ coefficients, where $d$ denotes the number of model parameters, and for degree 2 polynomials we have to jointly estimate $md(d+3)/2$ polynomial coefficients.
To achieve this using the plug-in principle, we must estimate covariance matrices containing $\mathcal{O}(m^2d^2)$ and $\mathcal{O}(m^2d^4)$ entries respectively.
Our results are therefore consistent with the finding that poor estimation of the polynomial coefficients can actually increase estimator variance \citep{Glasserman}.
It remains unclear how to develop control variates for these non-equilibrium estimators.

We exploited the ZV control variate scheme due to \cite{Mira} that permits the automatic construction of control variates for any statistical model in which the gradient of the log-likelihood (and the log-prior) are available.
More generally, we envisage that for models where these gradients are unavailable in closed form, the use of numerical approximations could provide a successful strategy \citep{Calderhead2}.
Results on benchmark datasets demonstrate that CTI outperforms standard TI, but that the difference in performance is reduced when the likelihood function is strongly multi-modal.
A natural direction for further research is to explore whether alternative control variates are better suited to these challenging problems.

CTI clearly inherits the theoretical and methodological challenges that are associated with control variates more generally.
In particular ZV control variates are not parametrisation-invariant and it is unclear how to select an optimal variance-minimising parametrisation.
Pertinent to CTI in particular, the optimal coefficients $\bm{\phi}^*(t)$ will vary smoothly with (inverse) temperature $t$ (SFig. \ref{coefficients2}), yet the conventional plug-in approach to estimation treats each rung $t_i$ of the temperature ladder independently, leading to rough trajectories (SFig. \ref{coefficients1}).
It would therefore be interesting to design an information sharing scheme that jointly estimates all coefficients.

The development of low-cost computational approaches to Bayesian model comparison is necessary for the widespread adoption of Bayesian methodology in hypothesis-driven research.
The extension of control variate strategies to this important setting offers a promising route towards achieving this goal.

\appendix

\section{Quadrature for TI} \label{A quadrature}

Implementations of TI employ quadrature to approximate the one dimensional integral in Eqn. \ref{ML}.
\cite{Friel} originally employed a simple trapezoidal rule whereby the (inverse) temperature domain $t \in [0,1]$ was partitioned using $0 = t_0 < t_1 < \dots < t_m = 1$ and the (log-)evidence was approximated by
\begin{eqnarray}
\log(p(\bm{y})) & \approx & \sum_{i = 0}^{m-1} \frac{(t_{i+1}-t_i)}{2} [\mathbb{E}_{\bm{\theta}|\bm{y},t_i}  \log(p(\bm{y}|\bm{\theta})) + \mathbb{E}_{\bm{\theta}|\bm{y},t_{i+1}}  \log(p(\bm{y}|\bm{\theta}))]. \label{quadr1}
\end{eqnarray}
The use of quadrature introduces bias into the resulting estimator. 
To reduce this quadrature error and thus the estimator bias, \cite{Friel2} proposed the second order correction term
\begin{eqnarray}
\sum_{i = 0}^{m-1} \frac{(t_{i+1}-t_i)^2}{12} [\mathbb{V}_{\bm{\theta}|\bm{y},t_{i+1}}  \log(p(\bm{y}|\bm{\theta})) - \mathbb{V}_{\bm{\theta}|\bm{y},t_i}  \log(p(\bm{y}|\bm{\theta}))]
\label{2nd quad}
\end{eqnarray}
that is subtracted from Eqn. \ref{quadr1}.
Here $\mathbb{V}_{\bm{\theta}|\bm{y},t} g(\bm{\theta})$ denotes the variance of the function $g(\bm{\theta})$, where $\bm{\theta}$ has distribution with density $p(\bm{\theta}|\bm{y},t)$.

\section{Asymptotic unbiasedness} \label{A ZV}

Propositions 1 and 2 of \cite{Mira} show that a sufficient conditions for asymptotic unbiasedness of ZV control variates, i.e. $\mathbb{E}_{\pi}[h(\bm{\theta})] = 0$, is that, in the case where $\bm{\Theta}$ is unbounded, $\lim_{r \nearrow \infty} \int_{\partial B_r} \pi \nabla P \cdot \bm{n} d\sigma = 0$ where $B_r \nearrow \bm{\Theta}$ is a sequence of bounded subsets and $\bm{n}$ denotes the versor orthogonal to the boundary $\partial B_r$.
This condition could be difficult to verify directly; below we contribute a sufficient condition $\bm{\Theta} = \mathbb{R}^d$ that is easily verified.
Consider a $d$-dimensional hypercube $B_r = \{\bm{\theta} : |\theta_i| \leq r/2\}$ with side length $r$ and surface area $2dr^{d-1}$ and let $k$ be the degree of the polynomial $P(\bm{\theta})$.
Then crude bounds give
$\int_{\partial B_r} \pi \nabla P \cdot \bm{n} d\sigma \leq \sup_{\bm{\theta} \in \partial B_r} |\pi(\bm{\theta}) \nabla P(\bm{\theta}) \cdot \bm{n}(\bm{\theta})| \times \int_{\partial B_r} d\sigma 
\leq \left[ \sup_{\|\bm{\theta}\|_1 \geq r} |\pi(\bm{\theta})| \right] \left[ \sup_{\bm{\theta} \in \partial B_r} \| \nabla P(\bm{\theta}) \|_1 \right] \times 2dr^{d-1}$.
Since $\sup_{\bm{\theta}\in\partial B_r}\|\nabla P(\bm{\theta})\| = \mathcal{O}(r^{k-1})$ it follows that a sufficient condition for unbiasedness of ZV is
\begin{eqnarray}
\left[ \sup_{\|\bm{\theta}\|_1 \geq r} \pi(\bm{\theta}) \right] r^{d+k-2} \rightarrow 0 \; \; \; \text{ as } r \rightarrow \infty.
\label{easy condition}
\end{eqnarray}
In practice this requires that the tails of the (unnormalised) density $\pi(\bm{\theta})$ vanish sufficiently quickly, with faster convergence required when higher degree polynomials are to be used.

\section{Second degree polynomials} \label{A QZV}

Second degree polynomials can be expressed as $P(\boldsymbol{\theta})=\bm{c}^{T}\boldsymbol{\theta}+ \frac{1}{2}\boldsymbol{\theta}^{T}\bm{B}\boldsymbol{\theta}$ where $\bm{c}$ is $d \times 1$ and $\bm{B}$ is $d \times d$.
This leads to ZV control variates of the form 
\begin{eqnarray}
h(\bm{\theta}) = -\frac{1}{2}\text{tr}(\bm{B})+ (\bm{c}+\bm{B}\boldsymbol{\theta})^{T}\bm{z}(\boldsymbol{\theta}), 
\label{eq:auxiliary:g:poly:quad}
\end{eqnarray}
where $\bm{c}$ and $\bm{B}$ denote the quadratic polynomial coefficients and $\text{tr}(\bm{B})$ is the trace of $\bm{B}$. 
We assume that $\bm{B}$ is symmetric, but this is not required in general.
Following \cite{Mira}, it is possible to rearrange the terms on the right hand side of Eqn. \ref{eq:auxiliary:g:poly:quad}
into the form $\bm{\phi}^T\bm{w}(\bm{\theta})$
where the column vectors $\bm{\phi}$, $\bm{w}(\boldsymbol{\theta})$
have $d(d+3)/2$ elements each, and are defined as
$\bm{\phi}:=[\bm{c}^T~\bm{d}^T~\bm{b}^T]^T$, where
  $\bm{d}$ is the diagonal of $\bm{B}$ and $\bm{b}$ is a column
  vector with $d(d-1)/2$ elements, whose element in the
  $(2d-j)(j-1)/2+(i-j)$ position is the lower
  diagonal $(i,j)$-th element of $\bm{B}$, and $\bm{w}:=[\bm{z}^T~\bm{u}^T~\bm{v}^T]^T$, where
  $\bm{u}:=\boldsymbol{\theta}\circ\bm{z}-\frac{1}{2}\bm{1}$, with
  $\circ,~\bm{1}$ denoting the Hadamard product and the unit vector
  respectively, while $\bm{v}$ is a column vector comprising
  $d(d-1)/2$ elements, whose element in the
  $(2d-j)(j-1)/2+(i-j)$ position equals
  $\theta_iz_j+\theta_jz_i,~j\in\{1,2,\dots,d\},~
  i\in\{2,3,\dots,d\},~j<i.$

The same derivation used to obtain Eqn. \ref{optimal phi} can be followed to deduce that optimal coefficients $\bm{\phi}^*$ in the case of second order polynomials are given by $\bm{\phi}^*= -\mathbb{V}_{\pi}^{-1}[\bm{w}(\boldsymbol{\theta})] \mathbb{E}_{\pi}[g(\boldsymbol{\theta})\bm{w}(\boldsymbol{\theta})]$.
Similarly the ZV strategy with degree 2 polynomials can be expected to reduce variance when a linear combination of the components of $\bm{w}(\bm{\theta})$ is highly correlated with the target function $g(\bm{\theta}) = \log p(\bm{y}|\bm{\theta})$.

\section{Formulae for Bayesian linear regression} 

\subsection{Known precision} \label{lin known formulae}

The power posterior follows $\bm{\beta}|\bm{y},t \sim N(\bm{\mu}(t),\bm{\Sigma}(t))$ where $\bm{\mu}(t) = \frac{t}{\sigma^2} \bm{\Sigma}(t) \bm{X}^T\bm{y}$, $\bm{\Sigma}(t)^{-1} = \frac{t}{\sigma^2}\bm{X}^T\bm{X} + \frac{1}{\zeta^2}\bm{I}$, whilst the integrand $\mathbb{E}_{\bm{\beta}|\bm{y},t}[\log p(\bm{y}|\bm{\beta},\sigma)]$ has the closed-form expression
$- \frac{n}{2} \log(2 \pi \sigma^2) -\frac{1}{2\sigma^2} (\bm{y} - \bm{X}\bm{\mu}(t))^T(\bm{y} - \bm{X}\bm{\mu}(t)) - \frac{1}{2\sigma^2} \text{tr}(\bm{X}^T\bm{X} \bm{\Sigma}(t))$ and the model evidence is
\begin{eqnarray}
p(\bm{y}) = \frac{1}{(2\pi)^{n/2}|\bm{\Omega}|^{1/2}} \exp\left\{-\frac{1}{2}\bm{y}^T\bm{\Omega}^{-1} \bm{y}\right\}
\end{eqnarray}
where $\bm{\Omega} = \sigma^2\bm{I} + \zeta^2\bm{X}\bm{X}^T$.

\subsection{Unknown precision} \label{lin unknown formulae}

Using the transformation $\tau \mapsto \eta = \log(\tau)$ we can ensure that the posterior $p(\bm{\theta}|\bm{y},t,m)$ is defined on $\mathbb{R}^d$ and has exponential tails so that, by Eqn. \ref{easy condition}, the unbiasedness condition is satisfied. 
For Model 1 we have $z_1 = -\frac{1}{2} t e^\eta\left( \sum_i y_i - \alpha - \beta\bar{x}_i \right) + \frac{1}{2} e^\eta r_0 (\alpha - 3000)$, $z_2 = -\frac{1}{2} t e^\eta \left( \sum_i (y_i - \alpha - \beta\bar{x}_i)\bar{x}_i \right) + \frac{1}{2} e^\eta s_0 (\beta-185)$ and $z_3 = -\frac{nt}{4} + \frac{t e^\eta}{4}\left( \sum_i (y_i - \alpha - \beta\bar{x}_i)^2 \right) - \frac{1+a_0}{2} + \frac{e^\eta}{2}\left[ b_0 + \frac{r_0}{2} (\alpha-3000)^2 + \frac{s_0}{2} (\beta - 185)^2 \right]$, where the components are ordered with respect to $\bm{\theta} = (\alpha,\beta,\eta)$.

Write $\bm{X}$ for the design matrix with $i$th row $[1,\bar{x}_i]$.
The model evidence, that is the object we wish to estimate, is given by
\begin{eqnarray}
p(\bm{y}) = \frac{b_0^{a_0}}{(2\pi)^{n/2}} \sqrt{\frac{|Q_0|}{|Q_n|}} \frac{\Gamma(a_n)}{\Gamma(a_0)} \left\{ b_0 + \frac{1}{2}\left[ \bm{y}'\bm{y} - \bm{B}_n^T \bm{Q}_n \bm{B}_n + \bm{B}_0^T \bm{Q}_0 \bm{B}_0 \right] \right\}^{-a_n}
\end{eqnarray}
where $a_n = a_0 + \frac{n}{2}$, $\bm{Q}_n = \bm{Q}_0 + \bm{X}^T\bm{X}$ and $\bm{B}_n = \bm{Q}_n^{-1}(\bm{X}^T\bm{y} + \bm{Q}_0 \bm{B}_0)$.
Derivations for Model 2 are analogous.

\section{Formulae for Goodwin Oscillator} \label{ap ode}

The Goodwin oscillator with $g$ species is given by
\begin{eqnarray}
\frac{dx_1}{ds} & = & \frac{a_1}{1+a_2x_g^\rho} - \alpha x_1 \label{Goodwin} \\
\frac{dx_2}{ds} & = & k_1x_1 - \alpha x_2 \nonumber \\
& \vdots & \nonumber \\
\frac{dx_g}{ds} & = & k_{g-1}x_{g-1} - \alpha x_g. \nonumber
\end{eqnarray}
Here $x_1$ represents the concentration of mRNA for a target gene and $x_2$ represents its corresponding protein product.
Additional variables $x_3,\dots,x_g$ represent intermediate protein species that facilitate a cascade of enzymatic activation that ultimately leads to a negative feedback, via $x_g$, on the rate at which mRNA is transcribed.
The solution $\bm{x}(s;\bm{\theta},\bm{x}_0)$ of this dynamical system depends upon synthesis rate constants $a_1$, $k_1,\dots,k_{g-1}$ and degradation rate constants $a_2$, $\alpha$.
The Goodwin oscillator permits oscillatory solutions only when $\rho > 8$.
Following \cite{Calderhead} we therefore set $\rho = 10$ as a fixed parameter. 
A $g$-variable Goodwin model as described above therefore has $g+2$ uncertain parameters $(a_1,a_2,k_1,\dots,k_{g-1},\alpha)$. 
The Goodwin oscillator does not permit a closed form solution, meaning that each evaluation of the likelihood function requires the numerical integration of the system in Eqn. \ref{Goodwin}.
Due to the substantive computational challenges associated with model comparison in this setting, we considered only 10 independent runs of population MCMC, each using only $N=1,000$ iterations.

\begin{figure*}[h]
\centering
\includegraphics[width = .8\textwidth]{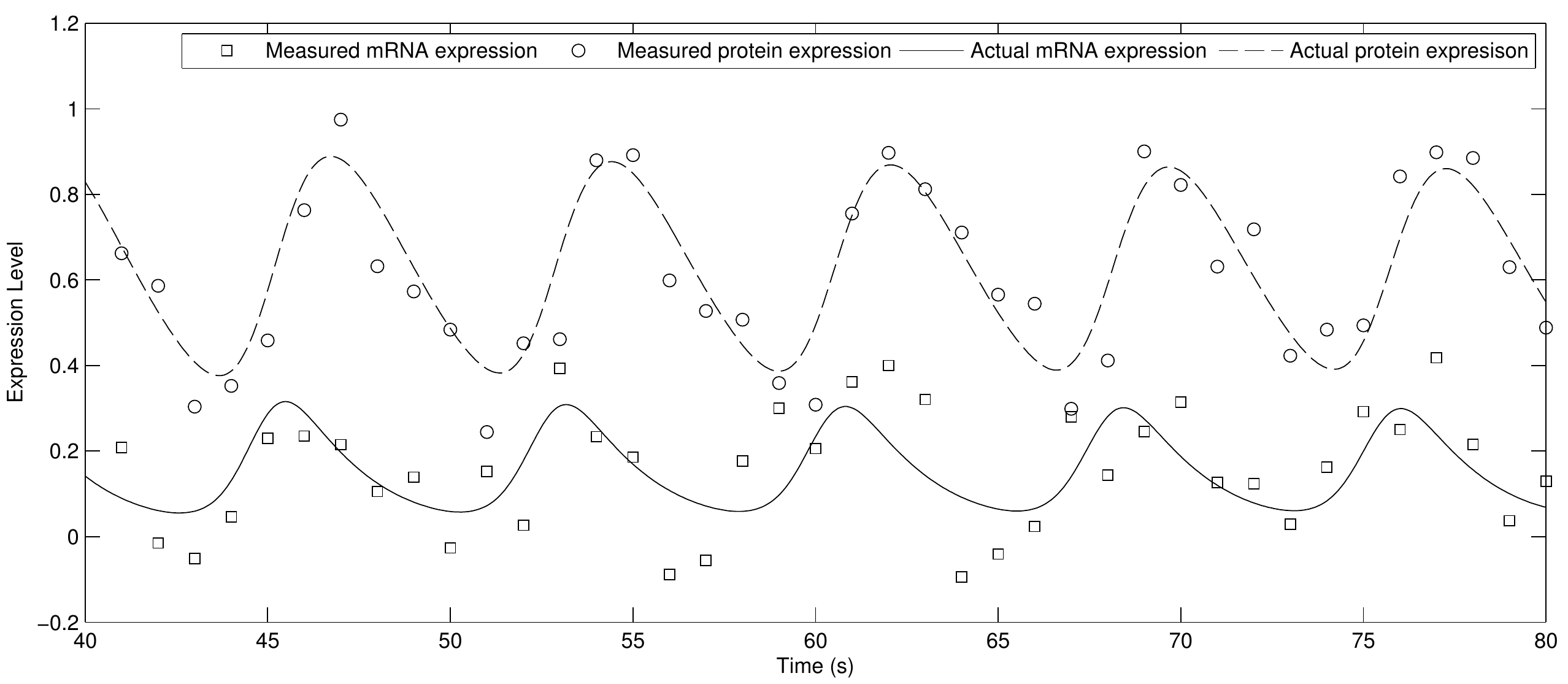}
\caption{Nonlinear ODEs: Data generated from the Goodwin oscillator based on $g=3$ species demonstrates characteristic oscillatory behaviour.}
\label{oscill}
\end{figure*}

We consider a realistic setting where only mRNA and protein product are observed, corresponding to $\bm{x}_a= [x_1,x_2]$. 
We assume $\bm{x}_0 = [0,\dots,0]$ and $\sigma = 0.1$ are both known and take sampling times to be $s = 41,\dots,80$.
Parameters were assigned independent $\Gamma(2,1)$ prior distributions. 
We generated data using $a_1 = 1$, $a_2 = 3$, $k_1 = 2$, $k_2,\dots,k_{g-1} = 1$, $\alpha = 0.5$, which produce oscillatory dynamics that do not depend heavily upon initial conditions (SFig. \ref{oscill}).

In practice we work with the log-transformed parameters $\bm{\theta}$.
In particular this allows us to verify that ZV methods are valid, since the tails of $p(\bm{\theta})$ vanish exponentially quickly.
Sensitivities $S_{j,k}^i$, defined in the main text, satisfy 
\begin{eqnarray}
\dot{S}_{j,k}^i = \frac{\partial f_k}{\partial \theta_i} + \sum_l \frac{\partial f_k}{\partial x_l} S_{j,l}^i \label{sense ode}
\end{eqnarray}
where $\frac{\partial x_k}{\partial\theta_i} = 0$ at $s = 0$.
Eqn. \ref{sense ode} provides a route to compute the sensitivities numerically, when they cannot be obtained analytically, by augmenting the state vector of the dynamical system to include the $S_{j,k}^i$.

\FloatBarrier

\newpage
\section*{Supplement}

\subsection*{Proof of exactness}

In this section we prove that CTI (degree 2) is exact (up to quadrature error) for the Bayesian linear regression model with known precision.

We have from Eqn. \ref{R} that the minimum variance ratio is given by
\begin{eqnarray}
R = 1 - \max_{\bm{\phi}} \text{Corr}_{\bm{\beta}|\bm{y},t}[g(\bm{\beta}),\bm{\phi}^T\bm{w}(\bm{\beta})].
\end{eqnarray}
Plugging in the expression of Eqn. \ref{eq:auxiliary:g:poly:quad} for $\bm{w}(\bm{\beta})$ we obtain
\begin{eqnarray}
R = 1 - \max_{\bm{B}, \bm{c}} \text{Corr}_{\bm{\beta}|\bm{y},t}[g(\bm{\beta}),-\frac{1}{2} \text{tr}(\bm{B}) + (\bm{c} + \bm{B}\bm{\beta})^T \bm{z}(\bm{\beta})] \label{eqblaah}
\end{eqnarray}
where the maximum is taken over all symmetric matrices $\bm{B}$ and real vectors $\bm{c}$.

Write $\stackrel{+C}{=}$ whenever two quantities are equal up to an additive constant not depending upon $\bm{\beta}$; since $\text{Corr}_{\bm{\beta}|\bm{y},t}[W,X] = \text{Corr}_{\bm{\beta}|\bm{y},t}[Y,Z]$ whenever $W \stackrel{+C}{=} Y$ and $X \stackrel{+C}{=} Z$, we need only work up to this equivalence.
We now claim that $\bm{z}(\bm{\beta})$ can be replaced with any transformation $\bm{z} \mapsto \bm{f} + \bm{E}\bm{z}$ in Eqn. \ref{eqblaah}, where we require that $\bm{E}$ is symmetric and invertible.
Indeed 
\begin{eqnarray}
(\bm{c}+\bm{B}\bm{\beta})^T(\bm{f} + \bm{E}\bm{z}(\bm{\beta})) \stackrel{+C}{=} (\bm{c}' + \bm{B}' \bm{\beta})^T \bm{z}(\bm{\beta}) + \bm{f}^T\bm{E}\bm{\beta} 
\label{comb1}
\end{eqnarray}
where $\bm{c}' = \bm{E}^T\bm{c}$, $\bm{B}' = \bm{E}^T\bm{B}$ (which is symmetric).
Moreover, from the definition of the control variates (Eqn. \ref{CVs1}) we have that $\bm{\beta} = 2 \bm{\Sigma}(t)[\bm{z}(\bm{\beta}) + \frac{t}{2\sigma^2}\bm{X}^T\bm{y}]$ and hence
\begin{eqnarray}
\bm{f}^T\bm{E}\bm{\beta}  \stackrel{+C}{=}  (\bm{c}'')^T \bm{z}(\bm{\beta})
\label{comb2}
\end{eqnarray}
where $\bm{c}'' = 2 \bm{f}^T\bm{E} \bm{\Sigma}(t)$.
Combining Eqns. \ref{comb1} and \ref{comb2} we have that
\begin{eqnarray}
(\bm{c}+\bm{B}\bm{\beta})^T(\bm{f} + \bm{E}\bm{z}(\bm{\beta})) \stackrel{+C}{=} (\bm{c}''' + \bm{B}' \bm{\beta})^T \bm{z}(\bm{\beta}) 
\end{eqnarray}
where $\bm{c}''' = \bm{c}'' + \bm{c}''$.
Recalling that correlation is invariant to the addition of constant terms, we have shown that 
\begin{eqnarray}
R \leq 1 - \max_{\bm{B}, \bm{c}} \text{Corr}_{\bm{\beta}|\bm{y},t}[g(\bm{\beta}),-\frac{1}{2} \text{tr}(\bm{B}) + (\bm{c} + \bm{B}\bm{\beta})^T (\bm{f} + \bm{E}\bm{z}(\bm{\beta}))].
\end{eqnarray}
In fact this equation is an equality, since the affine transformation is invertible and hence we can apply the same argument using the inverse transform.

Now $g(\bm{\beta}) \stackrel{+C}{=} (\bm{\beta}-\bm{m})^T \bm{S}^{-1} (\bm{\beta}-\bm{m})$ where $\bm{S} = (\bm{X}^T\bm{X}/\sigma^2)^{-1}$, $\bm{m} = \bm{S} \bm{X}^T \bm{y} / \sigma^2$.
Taking the specific choices $\bm{B} = \bm{S}^{-1}$ (which is symmetric), $\bm{c} = - \bm{S}^{-1}\bm{m}$, $\bm{f} = \frac{t}{\sigma^2} \bm{\Sigma}(t) \bm{X}^T \bm{y} - \bm{m}$ and $\bm{E} = 2 \bm{\Sigma}(t)$ (which is symmetric and invertible) we have
\begin{eqnarray}
R \leq 1 - \text{Corr}_{\bm{\beta}|\bm{y},t}[(\bm{\beta}-\bm{m})^T \bm{S}^{-1} (\bm{\beta}-\bm{m}),(\bm{\beta}-\bm{m})^T \bm{S}^{-1} (\bm{\beta}-\bm{m})] = 1 -1  = 0
\end{eqnarray}
which demonstrates that $R = 0$ and CTI (degree 2) is exact.

\subsection*{Manifold Metropolis-Adjusted Langevin Algorithm}

mMALA is a differential geometric MCMC scheme that, for power posteriors, requires that we have access to the metric tensor 
\begin{eqnarray}
\bm{G}(\bm{\theta}|t) = - \mathbb{E}_{\bm{y}|\bm{\theta}} \frac{\partial^2}{\partial\bm{\theta}^2} \log p(\bm{y},\bm{\theta}|t).
\end{eqnarray}
At current state $\bm{\theta}_n^{(i)}$ and for (inverse) temperature $t_i$ the ``simplified'' mMALA proposal follows from a discretised Langevin diffusion
\begin{eqnarray}
\bm{\theta}^*|\bm{\theta}_n^{(i)},\bm{y},t_i \sim N\left(\bm{\theta}_n^{(i)} + \frac{\epsilon^2}{2}  \bm{G}^{-1}(\bm{\theta}_n^{(i)}|\bm{y},t_i) \nabla_{\bm{\theta}} \log[p(\bm{y},\bm{\theta}_n^{(i)}|t_i)] , \epsilon^2 \bm{G}^{-1}(\bm{\theta}_n^{(i)}|\bm{y},t_i)\right)
\end{eqnarray}
that assumes constant curvature of the manifold.
The proposal $\bm{\theta}^*$ is then accepted as the next state $\bm{\theta}_{n+1}^{(i)}$ according to the Metropolis-Hastings ratio (else $\bm{\theta}_{n+1}^{(i)} = \bm{\theta}_n^{(i)}$).
For all applications in this paper we discarded the first $10\%$ of samples as burn-in and then retained the remaining $N$ samples for use.

The metric tensors for each of the applications considered in the Main Text are provided below:

{\bf Bayesian linear regression, known precision.}
\begin{eqnarray}
\bm{G}(\bm{\beta}|t) = \frac{t}{\sigma^2}\bm{X}^T\bm{X} - \frac{1}{\zeta^2} \bm{I}_{d \times d}
\end{eqnarray}
{\bf Bayesian linear regression, unknown precision (Radiata Pine).}
\begin{eqnarray}
\bm{G}(\bm{\theta}|t) = \left[ \begin{array}{ccc}  
e^\eta (nt+r_0) & 0 & e^\eta r_0 (\alpha-3000) \\
0 & e^\eta \left(s_0 + t \sum_i \bar{x}_i^2 \right) & e^\eta s_0 (\beta-185) \\
e^\eta r_0 (\alpha-3000) & e^\eta s_0 (\beta-185) & \frac{tn}{2} + e^\eta \left( b_0 + \frac{r_0}{2} (\alpha-3000)^2 + \frac{s_0}{2} (\beta-185)^2 \right)
\end{array} \right]
\end{eqnarray}
{\bf Bayesian logistic regression (Pima Indians).}
\begin{eqnarray}
G_{j,k}(\bm{\beta}|t) = -t \sum_i p_i(1-p_i)x_{i,j}x_{i,k} + \tau\delta_{j,k}
\end{eqnarray}
{\bf Bayesian inference for nonlinear ODEs (Goodwin Oscillator).}
\begin{eqnarray}
G_{i,l}(\bm{\theta}|t) = \delta_{i,l}\exp(\theta_i) + \frac{t}{\sigma^2} \sum_j [\bm{S}_{j,\bullet}^i] [\bm{S}_{j,\bullet}^l]^T
\end{eqnarray}

\subsection*{The controlled (equilibrated) annealed importance sampler} 

Annealed importance sampling (AIS) was proposed by \cite{Neal} as an extension of bridge sampling that improves mixing in parameter space by introducing multiple intermediate densities. 
In brief, AIS proceeds by producing samples $\bm{\theta}^{(0)}, \dots , \bm{\theta}^{m-1}$ as follows:
$\bm{\theta}^{(0)} \sim p(\bm{\theta})$.
Then $\bm{\theta}^{(j)} \sim T_j(\bm{\theta}^{(j-1)})$ in sequence for $j = 1,\dots,m-1$ where $T_j$ is a Markov transition kernel that targets the distribution $\bm{\theta}|\bm{y},t=t_j$.
Let $f(\bm{\theta}|\bm{y},t) = p(\bm{\theta}|\bm{y})^t p(\bm{\theta})$ so that $p(\bm{\theta}|\bm{y},t) = f(\bm{\theta}|\bm{y},t) / \mathcal{Z}_t(\bm{y})$.
Define
\begin{eqnarray}
w = \frac{f(\bm{\theta}^{(0)}|\bm{y},t_1)}{f(\bm{\theta}^{(0)}|\bm{y},t_0)} \cdot \frac{f(\bm{\theta}^{(1)}|\bm{y},t_2)}{f(\bm{\theta}^{(1)}|\bm{y},t_1)} \dots \frac{f(\bm{\theta}^{(m-1)}|\bm{y},t_m)}{f(\bm{\theta}^{(m-1)}|\bm{y},t_{m-1})}.
\end{eqnarray}
Then it is shown in \cite{Neal} that 
\begin{eqnarray}
\mathbb{E}_{(\bm{\theta}^{(0)}, \dots , \bm{\theta}^{(m-1)}) \sim G}[w] = \frac{\mathcal{Z}_1}{\mathcal{Z}_0} \cdot \frac{\mathcal{Z}_2}{\mathcal{Z}_1} \dots \frac{\mathcal{Z}_m}{\mathcal{Z}_{m-1}} = \frac{\mathcal{Z}_m}{\mathcal{Z}_0} = p(\bm{y})
\end{eqnarray}
where the expectation is over the generative process $G$ described above. Note that this is precisely $m$ versions of bridge sampling, each targeting one of the ratios in the above equation.

AIS is a non-equilibrium estimator, in the sense that the marginal distribution of $\bm{\theta}^{(i)}$ need not be the same as the distribution $\bm{\theta}|\bm{y},t_i$, and is therefore not directly amenable to ZV control variates.
In order to transform AIS into an equilibrium estimator we need to consider jointly sampling all the $\bm{\theta}^{(i)}$.
Specifically, we exploit the fact that
\begin{eqnarray}
\mathbb{E}_{(\bm{\theta}^{(0)}, \dots , \bm{\theta}^{(m-1)}) \sim G}[w] = \mathbb{E}_{\substack{\bm{\theta}^{(i)}|\bm{y},t_i \\ 0 \leq i \leq m-1}}[w].
\end{eqnarray}
Estimation in the equilibrated AIS therefore requires a collection of samples $\bm{\theta}^{(j)} \sim \bm{\theta}|\bm{y},t_j$ that can be obtained using (converged) MCMC.
In this paper we generated these samples using population MCMC \citep{Jasra}; for fair comparison we used the same samples that were the basis for TI experiments.

Rewriting $w$ as in \cite{Vyshemirsky} we obtain
\begin{eqnarray}
p(\bm{y}) = \mathbb{E}_{\substack{\bm{\theta}^{(i)}|\bm{y},t_i \\ 0 \leq i \leq m-1}} \left[ \exp\left( \sum_{i=0}^{m-1} (t_{i+1}-t_i) \log(p(\bm{y}|\bm{\theta}^{(i)})) \right) \right].
\label{AIS}
\end{eqnarray}
Since a Monte Carlo estimate based on Eqn.~\ref{AIS} will be unbiased, we need simply choose the temperature ladder sufficiently fine that our acceptance rates indicate good mixing.
In experiments below, for fairness of comparison, the same temperature ladder was used for (C)AIS as for (C)TI.
This controls the amount of information present in the samples $\bm{\theta}_n^{(i)}$ and allows the samples from the same run of population MCMC to be used for all estimators.

The Monte Carlo expectation for equilibrated AIS is taken over all $\bm{\theta}^{(0:m-1)} = \{\bm{\theta}^{(i)}\}_{i=0}^m$ simultaneously; we therefore base ZV control variates on
\begin{eqnarray}
\bm{z}(\bm{\theta}^{(0:m-1)}|\bm{y},t_{0:m-1}) = -\frac{1}{2} \nabla_{\bm{\theta}^{(0:m-1)}} \log \left[ \prod_{i=0}^{m-1} p(\bm{\theta}^{(i)}|\bm{y},t_i) \right]
\end{eqnarray}
so that $\bm{z}(\bm{\theta}^{(0:m-1)}|\bm{y},t_{0:m-1})$ has a block structure whose components are given by Eqn.~\ref{Z path}.
Then ZV control variates are given by
\begin{eqnarray}
h(\bm{\theta}^{(0:m-1)}|\bm{y},t_{0:m-1}) & = & 
-\frac{1}{2} \Delta_{\bm{\theta}^{(0:m-1)}}[P(\bm{\theta}^{(0:m-1)}|\bm{\phi}(\bm{y},t_{0:m-1}))]  \\
& & +  \nabla_{\bm{\theta}^{(0:m-1)}}[P(\bm{\theta}^{(0:m-1)}|\bm{\phi}(\bm{y},t_{0:m-1}))] \cdot \bm{z}(\bm{\theta}^{(0:m-1)}|\bm{y},t_{0:m-1}). \nonumber
\end{eqnarray}

The CAIS estimator is defined by the identity
\begin{eqnarray}
p(\bm{y}) = \mathbb{E}_{\substack{\bm{\theta}^{(i)}|\bm{y},t_i \\ 0 \leq i \leq m-1}} \left[ \exp\left( \sum_{i=0}^{m-1} (t_{i+1}-t_i) \log(p(\bm{y}|\bm{\theta}^{(i)})) \right) + h(\bm{\theta}^{(0:m-1)}|\bm{y},t_{0:m-1}) \right].
\label{bridge improve}
\end{eqnarray}
When coefficients $\bm{\phi}(\bm{y},t_{0:m-1})$ are chosen optimally, the Monte Carlo estimator of Eqn. \ref{bridge improve} will have variance that is, in the worst case, no larger than the variance of the standard AIS estimator.
In practice, polynomial coefficients are estimated using the plug-in approach of Eqn. \ref{plug}, taking $g(\bm{\theta}) =  \exp\left( \sum_{i=0}^{m-1} (t_{i+1}-t_i) \log(p(\bm{y}|\bm{\theta}^{(i)})) \right)$.

As discussed in the main text, the plug-in approach typically fails due to the high-dimensionality of the covariance matrices that must be estimated.
In addition, implementation of CAIS is complicated due to the requirement that the integrand of Eqn. \ref{bridge improve} must remain positive; this further detracts from the suitability of CAIS.

\FloatBarrier
\subsection*{Additional figures}

\begin{figure}[h]
\centering
\begin{subfigure}[]{.45\textwidth}
\includegraphics[clip,trim = 0cm 0cm 13cm 0.6cm,width = \textwidth]{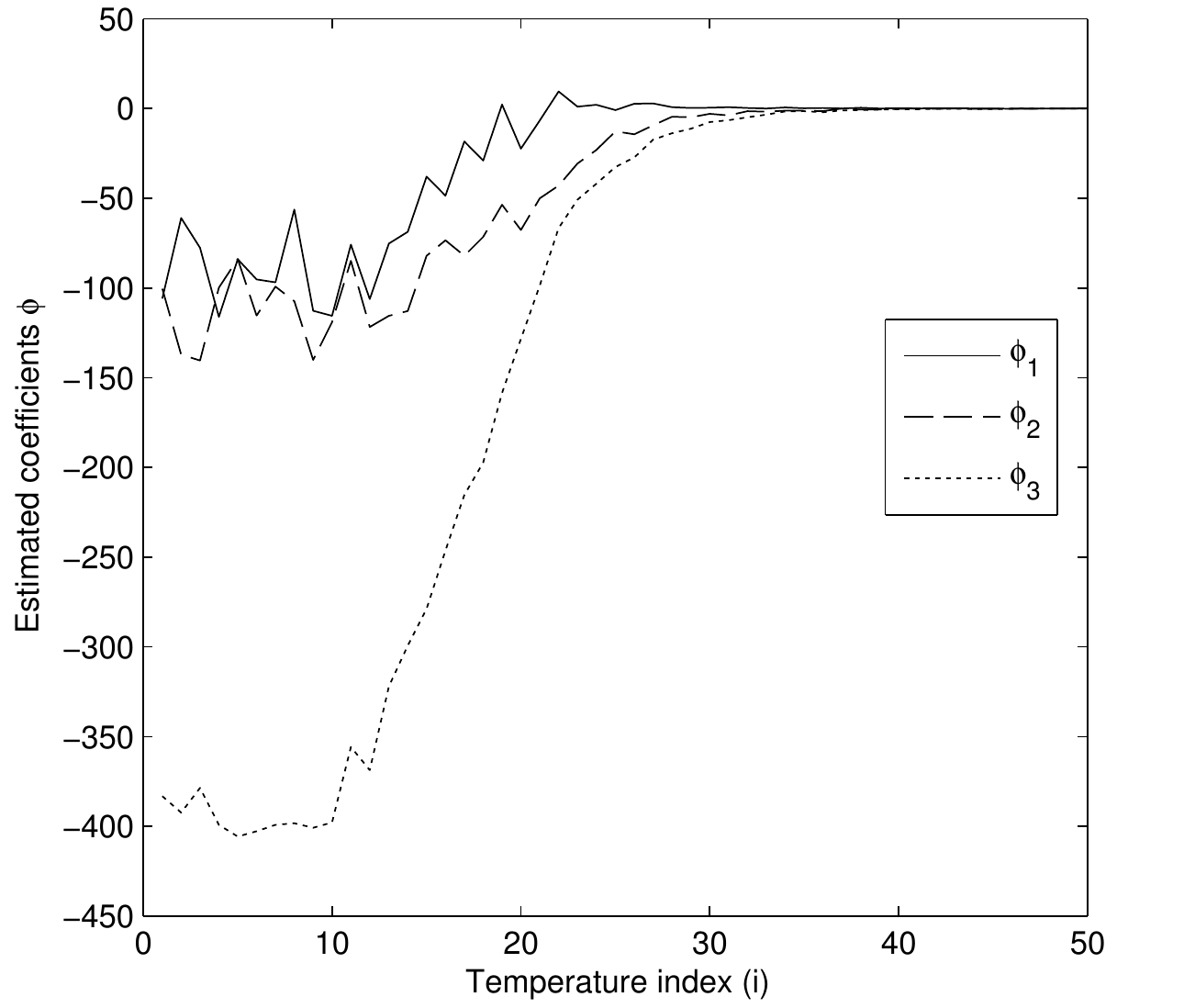} 
\caption{Degree 1}
\label{coefficients1}
\end{subfigure}
\begin{subfigure}[]{.45\textwidth}
\includegraphics[clip,trim = 13cm 0cm 0cm 0.6cm,width = \textwidth]{Figures/est_coeff.pdf} 
\caption{Degree 2}
\label{coefficients2}
\end{subfigure}
\caption{Estimated polynomial coefficients $\bm{\phi}^*(t_i)$ for ZV control variates. 
(a) Degree 1 polynomials. (b) Degree 2 polynomials.
[Here we show one particular realisation corresponding to one run of population MCMC. It can be seen that, for degree 2 polynomials, the plug-in estimate for coefficients is deterministic.
The x-axis records the index $i$ corresponding to (inverse) temperature $t_i = (i/50)^5$.]
}
\label{coefficients}
\end{figure}

\begin{figure}[h]
\centering
\begin{subfigure}[]{.3\textwidth}
\includegraphics[width = \textwidth,clip,trim = 0cm 0cm 14.8cm 0.65cm]{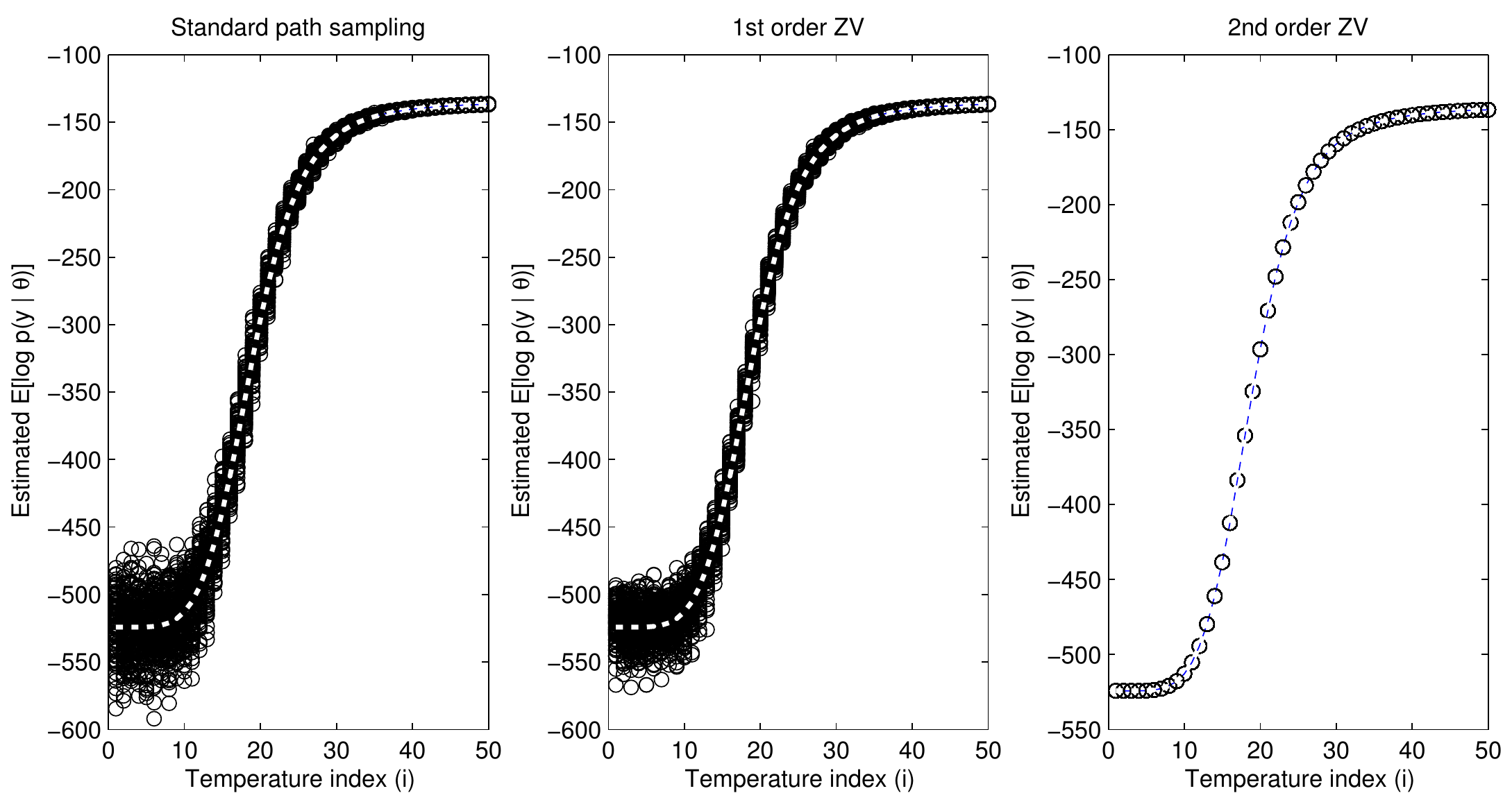} 
\caption{Degree 1}
\end{subfigure}
\begin{subfigure}[]{.3\textwidth}
\includegraphics[width = \textwidth,clip,trim = 7.4cm 0cm 7.4cm 0.65cm]{Figures/integrands.pdf} 
\caption{Degree 2}
\end{subfigure}
\begin{subfigure}[]{.3\textwidth}
\includegraphics[width = \textwidth,clip,trim = 14.8cm 0cm 0cm 0.65cm]{Figures/integrands.pdf} 
\caption{Degree 3}
\end{subfigure}
\caption{Estimates for the integrand $\mathbb{E}_{\bm{\beta}|\bm{y},t}[\log p(\bm{y}|\bm{\theta})]$, based on 100 independent runs of population MCMC with $N=1000$ samples and a quintic temperature ladder $t_i = (i/50)^5$. The dashed blue/white curve represents the true value of the integrand.
[Here we consider polynomial trial functions $P(\bm{\theta})$ of (a) degree 0 (i.e. standard TI), (b) degree 1 and (c) degree 2.
The x-axis records the index $i$ corresponding to (inverse) temperature $t_i = (i/50)^5$.]
}
\label{integrands}
\end{figure}

\begin{figure*}[h]
\centering
\begin{subfigure}[]{.6\textwidth}
\includegraphics[width = \textwidth]{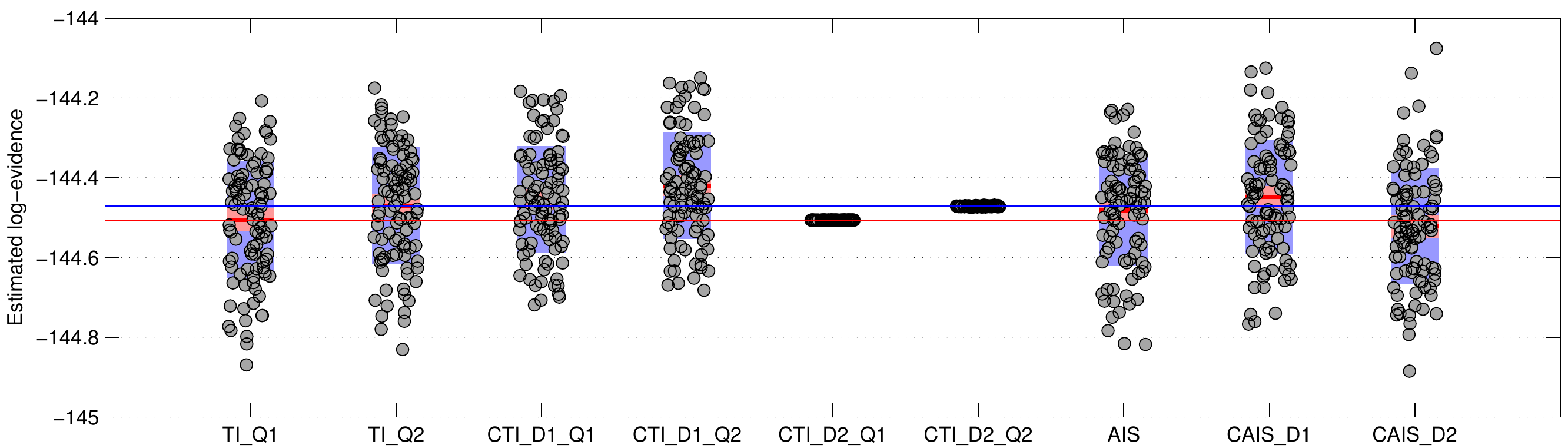}
\caption{Bayesian linear regression, known precision}
\label{var reduce}
\end{subfigure}

\begin{subfigure}[]{.6\textwidth}
\includegraphics[width = \textwidth]{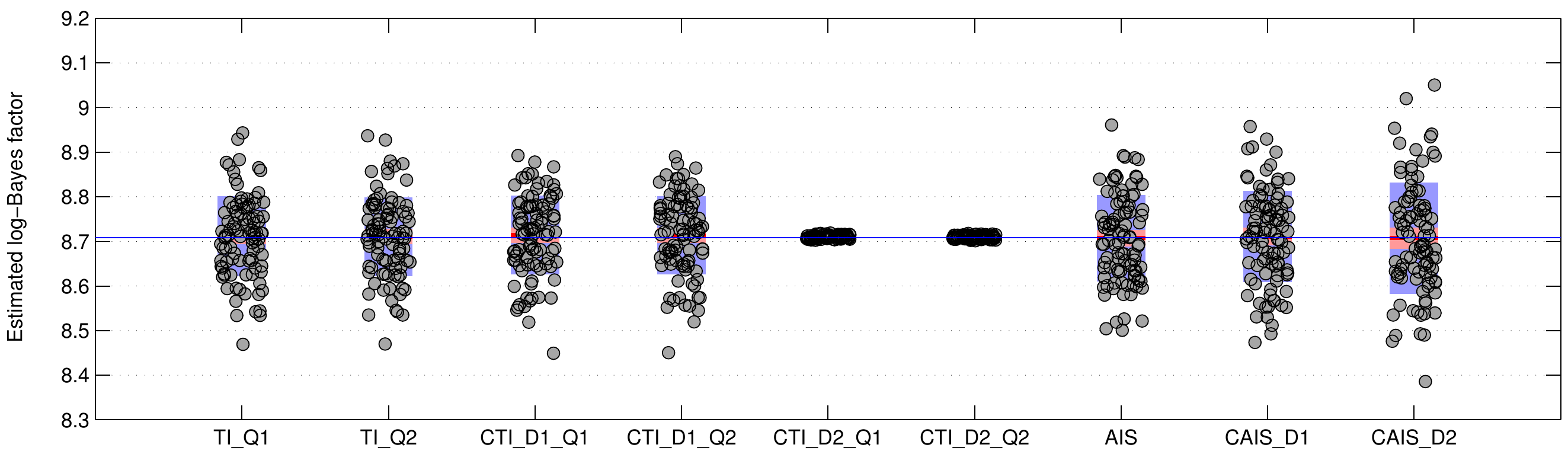}
\caption{Bayesian linear regression, unknown precision}
\label{app2 BF}
\end{subfigure}

\begin{subfigure}[]{.6\textwidth}
\includegraphics[width = \textwidth]{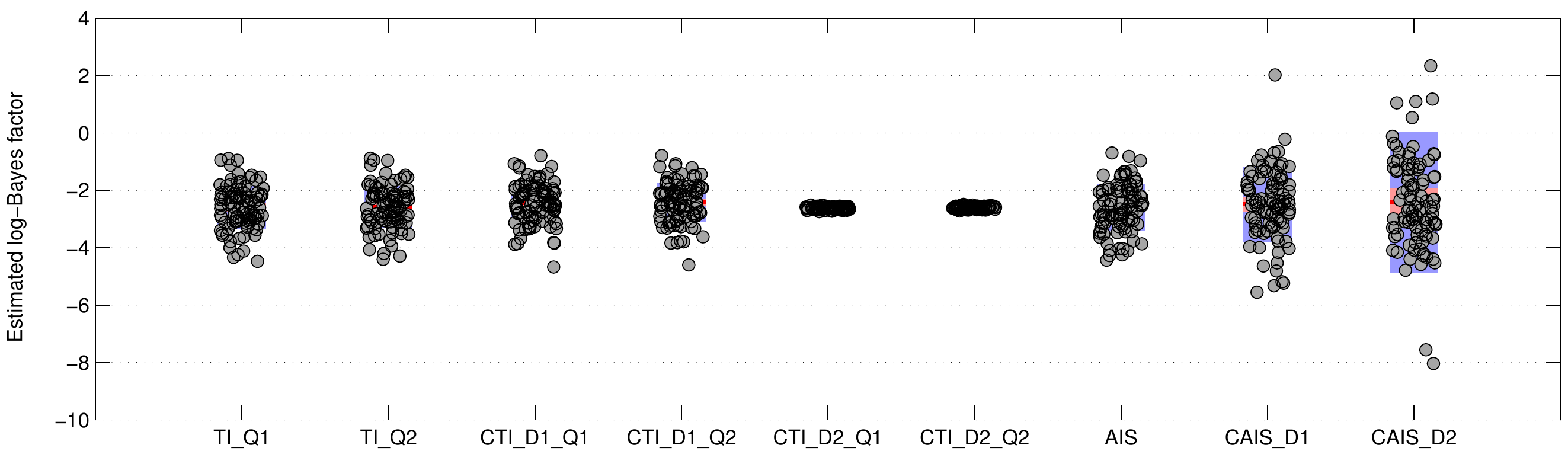}
\caption{Bayesian logistic regression (Radiata Pine)}
\label{BF logistic}
\end{subfigure}

\begin{subfigure}[]{.6\textwidth}
\includegraphics[width = \textwidth]{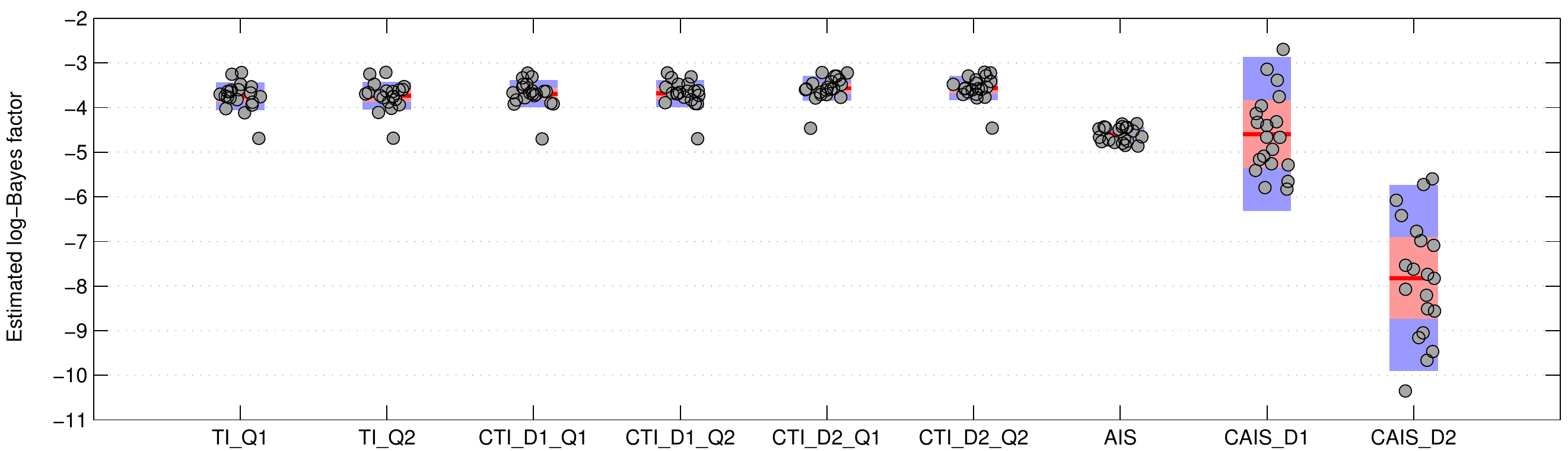}
\caption{Nonlinear ODEs (Goodwin Oscillator)}
\label{ode scatter}
\end{subfigure}

\caption{Estimates for evidence/Bayes factors.
(a) Bayesian linear regression, known precision: Estimates of log-evidence, based on 100 independent runs of population MCMC with $N=1000$ samples. 
The blue line shows the true log-evidence, whereas the red line displays the biased form of the log-evidence when first order quadrature error is taken into account.
(b) Radiata pine: Estimates of the log-Bayes factor of Model 2 in favour of Model 1, based on 100 independent runs of population MCMC with $N=1000$ samples. 
The blue line shows the true log-Bayes factor, which is $B_{12} = 8.7086$.
(c) Pima Indians: Estimates of the log-Bayes factor of Model 2 in favour of Model 1, based on 100 independent runs of population MCMC with $N=1000$ samples. 
(d) Goodwin oscillator: Estimates of the log-Bayes factor of Model 2 in favour of Model 1, based on 10 independent runs of population MCMC with $N=1000$ samples.
[TI = thermodynamic integration, CTI = controlled TI, AIS = annealed importance sampling, CAIS = controlled AIS, D1 = degree 1 polynomials, D2 = degree 2 polynomials, Q1 = first order quadrature, Q2 = second order quadrature.
Red error regions are used to display 95\% confidence intervals for the sample mean over all estimates, and blue error regions display the inter-quartile range for the estimates.]
}
\end{figure*}

\end{document}